\documentclass[10pt, journal]{IEEEtran}

\usepackage{cite}

\makeatletter
\def\ps@headings{%
\def\@oddhead{\mbox{}\scriptsize\rightmark \hfil \thepage}%
\def\@evenhead{\scriptsize\thepage \hfil \leftmark\mbox{}}%
\def\@oddfoot{}%
\def\@evenfoot{}}
\makeatother
\usepackage{multirow}
\usepackage{graphicx}
\usepackage{verbatim}
\usepackage{amsthm}
\usepackage{amsmath}
\usepackage{mathrsfs}
\usepackage{subfigure}
\usepackage{epstopdf}
\usepackage{caption}

\usepackage{longtable}
\usepackage{lipsum}
\usepackage{comment}
\usepackage{amsmath,bm}
\usepackage{booktabs}
\usepackage{color}
\usepackage{caption}
\usepackage{float} 
\usepackage{cite}
\usepackage{url}
\usepackage{booktabs}
\usepackage{amsmath,amssymb,amsfonts}
\usepackage{float}

\usepackage{graphicx}
\usepackage{algorithm}
\usepackage{algorithmic}
\usepackage{textcomp}
\usepackage{xcolor}
\usepackage{cuted} 

\UseRawInputEncoding

\begin{document}

\title{Matching-Driven Deep Reinforcement Learning for Energy-Efficient Transmission Parameter Allocation in Multi-Gateway LoRa Networks} 




\author{\IEEEauthorblockN{
		Ziqi~Lin,
		Xu~Zhang,
		Shimin~Gong,~\IEEEmembership{Member,~IEEE,}
		Lanhua~Li,
		Zhou~Su,~\IEEEmembership{Senior Member,~IEEE,} 
		and 		Bo~Gu,~\IEEEmembership{Member,~IEEE}
	}
	\thanks{This work was supported in part by the National Science Foundation of China (NSFC) under Grant U20A20175 and in part by the National Key R\&D Program of China under Grant 2020YFB1713800. \textit{(Corresponding author: Bo Gu.)}}
	\IEEEcompsocitemizethanks{
		\IEEEcompsocthanksitem B. Gu, S. Gong, L. Li, Z. Lin and X. Zhang are with the School of Intelligent Systems Engineering, Sun Yat-sen University, Shenzhen 518107, China, and with the Guangdong Provincial Key Laboratory of Fire Science and Intelligent Emergency Technology, Guangzhou 510006, China (E-mails: gubo@mail.sysu.edu.cn;  gong0012@e.ntu.edu.sg;
		lilh65@mail.sysu.edu.cn;
		linzq36@mail2.sysu.edu.cn; 
		zhangx526@mail2.sysu.edu.cn).
		\IEEEcompsocthanksitem Z. Su is with the School of Cyber Science and Engineering, Xi'an Jiaotong
		University, Xi'an 710049, China (E-mail: zhousu@ieee.org).
}}

\maketitle

\begin{abstract}
Long-range (LoRa) communication technology, distinguished by its low power consumption and long communication range, is widely used in the Internet of Things. Nevertheless, the LoRa MAC layer adopts pure ALOHA for medium access control, which may suffer from severe packet collisions as the network scale expands, consequently reducing the system energy efficiency (EE). To address this issue, it is critical to carefully allocate transmission parameters such as the channel (CH), transmission power (TP) and spreading factor (SF) to each end device (ED). Owing to the low duty cycle and sporadic traffic of LoRa networks, evaluating the system EE under various parameter settings proves to be time-consuming. Consequently, we propose an analytical model aimed at calculating the system EE while fully considering the impact of multiple gateways, duty cycling, quasi-orthogonal SFs and capture effects. On this basis, we investigate a joint CH, SF and TP allocation problem, with the objective of optimizing the system EE for uplink transmissions. Due to the NP-hard complexity of the problem, the optimization problem is decomposed into two subproblems: CH assignment and SF/TP assignment. First, a matching-based algorithm is introduced to address the CH assignment subproblem. Then, an attention-based multiagent reinforcement learning technique is employed to address the SF/TP assignment subproblem for EDs allocated to the same CH, which reduces the number of learning agents to achieve fast convergence. The simulation outcomes indicate that the proposed approach converges quickly under various parameter settings and obtains significantly better system EE than baseline algorithms.

\end{abstract}


\IEEEpeerreviewmaketitle

\section{INTRODUCTION}
Owing to their long communication range and low power consumption, low-power wide area networks (LPWANs) \cite{raza2017low} are extensively utilized in the development of autonomous wireless networks for various Internet of Things (IoT) applications, such as smart agriculture \cite{lu2020energy} and smart cities \cite{eletreby2017empowering}.
As typical LPWANs, long-range (LoRa) networks have attracted increasing attention from researchers in both industry and academia since they operate in licence-free ISM frequency bands \cite{sundaram2019survey}.

A LoRa network typically comprises numerous end devices (EDs), LoRa gateways (GWs), and a network server (NS). Data packets from EDs are collected by GWs and subsequently forwarded to the NS. In the MAC layer, the LoRa network adopts pure ALOHA to access the wireless medium \cite{LoRaWAN}. However, the randomness of pure ALOHA may lead to serious packet collisions with an increase in the number of EDs. Data retransmission caused by such collisions may escalate the energy consumption of EDs and consequently reduce the system energy efficiency (EE).

Different combinations of EDs' transmission parameters, such as the channel (CH), spreading factor (SF), and transmission power (TP), result in varied performance. For instance, increasing the TP and SF enhances the transmission's resilience to noise and increases the transmission range but concurrently decreases the data rate and increases the energy consumption.
Existing work on transmission parameter allocation for EDs in LoRa networks has focused mainly on single-GW scenarios \cite{qin2022channel, 9417579, sherazi2020energy}.
In this paper, we concentrate on optimizing the system EE of a multi-GW LoRa network through the allocation of the transmission parameters, which is challenging for the following reasons.

1) \textbf{Lack of an analytical model for calculating the EE:} LoRa EDs perform with low duty cycles and have limited data rates, which results in sporadic packet transmission. Therefore, it is time-consuming to evaluate the EE across various parameter settings in practice. An analytical model that can evaluate the EE in real time is therefore important for efficiently allocating the transmission parameters. However, most existing analytical models consider only the single-GW scenario \cite{li2020dylora, mahmood2018scalability}. In multi-GW scenarios, packets from an ED may experience independent path loss and can be received by any GW within the communication range. Therefore, existing analytical models for single-GW scenarios cannot be directly applied to multi-GW scenarios.

2) \textbf{Imperfect SF orthogonality:} Due to imperfect SF orthogonality, the interference resulting from concurrent transmission over the same CH with different SFs (i.e., inter-SF interference) is nonnegligible \cite{benkhelifa2021user}. Such inter-SF interference may increase the possibility of packet loss when the signal-to-interference ratio (SIR) is below a specific threshold \cite{8292748}. Since packets can be received by several GWs, different combinations of transmission parameters may result in various packet delivery rates (PDR) in different GWs, which further complicates EE optimization in multi-GW scenarios.

To address the abovementioned challenges, we design an analytical model to calculate the system EE within a multi-GW LoRa network.
Then, we investigate the problem of joint CH, SF and TP allocation to optimize the system EE. Due to the NP-hard complexity of the problem, we decompose the optimization problem into two subproblems: CH assignment and SF/TP assignment. A two-stage optimization framework, termed MMALoRa, which integrates matching techniques and multiagent deep reinforcement learning, is proposed to maximize the system EE. 
First, a matching-based algorithm is designed to solve the CH assignment subproblem.
Then, an attention-based multiagent reinforcement learning (MAAC) algorithm \cite{iqbal2019actor} is employed to address the SF/TP assignment problem.
The principal contributions of this paper are fourfold:
\begin{itemize}
	\item We propose an analytical model to calculate the system EE in multi-GW LoRa networks while fully considering the imperfect SF orthogonality and duty cycling. The simulation outcomes confirm that the proposed model can accurately evaluate the system EE of LoRa networks under different parameter settings.
	
	\item Leveraging the proposed analytical model, we formulate a joint CH/SF/TP assignment problem aimed at maximizing the uplink systems' EE. Then, we propose a two-stage optimization framework, namely MMALoRa, which combines matching theory and multiagent reinforcement learning to derive the optimal solution.
	
	\item CH assignment is formulated as a many-to-one matching problem and a low-complexity algorithm is developed to derive stable matching results between EDs and CHs. To achieve fast convergence, an MAAC-based algorithm is proposed to determine the SF/TP for each ED group assigned to the same CH to reduce the number of learning agents. In particular, the attention mechanism enables each ED to determine how much ``attention'' should be given to the parameter allocations of pertinent EDs with the goal of enhancing the system EE.
	
	\item Simulation results reveal that MMALoRa can achieve rapid convergence and that the EDs can learn appropriate transmission parameter allocation policies in multi-GW LoRa networks.
\end{itemize}

\section{RELATED WORK} 

Recently, various techniques have been employed to optimize the parameters of EDs in LoRa networks, aiming to enhance network performance.
LoRaWAN, the MAC layer protocol of LoRa networks, currently utilizes an adaptive data rate (ADR) \cite{garlisi2020capture} to automatically select the SF and TP for each ED to ensure that the signal-to-noise ratio (SNR) of the ED is higher than a demodulation floor.
However, Li $et$ $al.$ \cite{li2020dylora} demonstrated that the ADR approach tends to choose larger TPs and SFs to meet the minimum SNR required for demodulation, potentially resulting in an increase in energy consumption of as much as 103\%.

To improve the system EE of LoRa networks, many advanced resource allocation algorithms have been proposed \cite{li2020dylora, bor2017lora}.
Li $et$ $al.$ \cite{li2020dylora} designed DyLoRa to achieve optimal EE by dynamically adjusting the optimal transmission parameters for EDs.
Martin $et$ $al.$ \cite{bor2017lora} proposed a link quality detection mechanism (i.e., calculating lost/erroneous data packets over time) to quickly identify the optimal transmission parameters to improve system EE.
However, these methods rely on the assumption of perfectly orthogonal SFs, which is untenable in real LoRa networks.
Several studies have focused on parameter allocation in scenarios with imperfect SF orthogonality \cite{amichi2020joint, su2020energy}.
Amichi $et$ $al.$ \cite{amichi2020joint} investigated the issue of jointly optimizing SF and TP allocation to improve throughput fairness and reduce energy consumption.
Su $et$ $al.$ \cite{su2020energy} focused on uplink transmissions and proposed a joint user scheduling, SF allocation and TP assignment scheme to maximize the EE. However, the abovementioned methods require frequent information exchange between the GW and EDs, which is difficult to implement due to the constrained spectrum resources in LoRa networks.

Many studies have investigated parameter assignment within LoRa networks employing deep reinforcement learning (DRL).
Inaam $et$ $al.$ \cite{ilahi2020intelligent} introduced a centralized DRL-based scheme for parameter assignment in which the GW acts as a centralized node to adjust the parameters of each ED to maximize the PDR.
Qin $et$ $al.$ \cite{qin2022channel} focused on maximizing the system EE by training a deep Q-network based on the GW to generate CH and SF allocations and then solving a nonconvex optimization problem to determine the optimal TP.
Duc-Tuyen $et$ $al.$ \cite{ta2019lora} designed a multiarmed bandit-based learning algorithm to minimize the probability of collisions by appropriately allocating radio resources.

Different from the abovementioned works focused on single-GW scenarios, some researches \cite{zhao2022towards, liao2019adaptive, bouazizi2020spatiotemporal} have studied the parameter allocation problem in multi-GW LoRa networks.
Liao $et$ $al.$ \cite{liao2019adaptive} introduced a dynamic approach for selecting the GW and assigning SFs, which reduces the packet collision rate by equalizing the time-on-air (ToA) of each SF.
Yathreb $et$ $al.$ \cite{bouazizi2020spatiotemporal} presented a stochastic geometry-based approach for analyzing the uplink transmission performance in LoRa networks.
Gao $et$ $al.$ \cite{zhao2022towards} introduced EF-LoRa, a framework designed to achieve fair EE among EDs by allocating different parameters to various EDs.
However, the above investigations neglected the impact of capture effects and quasiorthogonality between SFs on the performance of LoRa networks; thus, their results do not represent real-world scenarios.

In this paper, we focus on a multi-GW LoRa network and propose a two-stage framework, MMALoRa, to optimize the system EE while fully considering imperfect SF orthogonality, capture effects, and duty cycling.
Since the NS has global knowledge of all EDs, MMALoRa can run on a high-performance NS. Thus, the learned experiences of each ED can be shared through the NS, thereby improving the efficiency of model training.

\section{SYSTEM MODEL}

We focus on uplink transmissions in a multi-GW LoRa network consisting of $N$ EDs, $K$ GWs, and one NS.
The sets of GWs and EDs are denoted as ${\cal K} = \{ 1,2,\cdots,K\} $ and ${\cal N} = \{ 1,2,\cdots,N\} $, respectively.
Specifically, the GWs receive and decode packets from the EDs deployed within their communication range and forward them to the NS.
The LoRa EDs communicate with the GW through multiple CHs and adopt SFs to share frequencies and time slots with EDs within the same CH. At most, each ED is allowed to access one CH and select one SF per transmission. The sets of SFs and CHs are expressed as ${\cal M} = \{ 7,8,\cdots,12\}$ and ${\cal C} = \{ 1,2,\cdots,C\} $, respectively.

\subsection{The Traffic Pattern of LoRa Networks}

We assume that the traffic pattern of the LoRa network follows a Poisson distribution \cite{toro2021modeling}. Therefore, an ED transmits $\kappa$ packets within a time interval $T$ with the following probability:
\begin{equation}
\mathbb{P}\left( {X = \kappa} \right) = \frac{{{{\left( {\lambda T} \right)}^\kappa}}}{{\kappa!}}{e^{ - \lambda T}}
\end{equation}
where $\lambda $ denotes the average sending rate.
Accordingly, the probability that the ED remains silent in time interval $T$ is $\mathbb{P}\left( {X = 0} \right) = e^{ - \lambda T}$. Then, the probability that the ED transmits at least one packet can be defined as:
\begin{equation}
h = 1 - \mathbb{P}\left( {X = 0} \right) = 1-{e^{ - \lambda T}}
\end{equation}

\begin{figure}[tb]
	\centering
	\includegraphics[width=75mm]{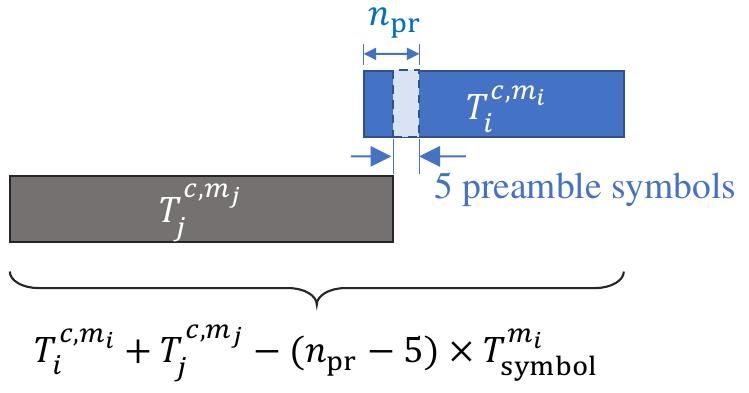}
	\caption{Packets from the target ED and interferer with interference time interval and preamble details.
	}\label{time2}
\end{figure}

As shown in Fig. \ref{time2}, Bor $et$ $al.$ \cite{bor2016lora} experimentally demonstrated that even if the preamble is affected by interference, the packet can still be decoded if the GW can correctly receive the last five symbols of the preamble. Therefore, to guarantee the successful reception of the packet from ED $i$ by the GW, the interfering ED $j$ that uses the same CH as ED $i$ (i.e., $c_i=c_j$) should not transmit during the following interference time interval:
\begin{equation}
\begin{aligned}
&{T_{ij}} = T_j^{c_j,m_j} + T_i^{c_i,m_i} - ({n_{\text{pr}}} - 5) \times T_{\text{symbol}}^{c_i, {m_i}}
\end{aligned}
\end{equation}
where $T_i^{c_i,m_i}$ and $T_j^{c_i,m_j}$ are the ToA for ED $i$ and interfering ED $j$, respectively, to transmit a packet when using SF $m_i$ and $m_j$, respectively. The ToA of the packets transmitted by ED $i$ is given by:
\begin{equation}
\begin{aligned}
&T_i^{c_i,m_i} = (n_\text{pr}+n_{\text{pl}})\times T_{\text{symbol}}^{c_i, {m_i}}\\
&=\left(20.25 + \max (\left\lceil {\frac{{8L - 4{m_i} + 28 + 16}}{{4({m_i} - 2DE)}}} \right\rceil CR ,0) \right)\times T_{\text{symbol}}^{c_i, {m_i}}
\end{aligned}
\end{equation}
where $n_{\text{pr}}$ and $n_{\text{pl}}$ represent the numbers of symbols of the preamble and the payload, respectively. $L$ represents the payload size of the packet, and $DE$ is the low-data-rate mode indicator, which is $DE=1$ if the mode is enable and $DE=0$ otherwise. The $CR$ value ranges from 5 to 8, representing a coding rate ranging from 4/5 to 4/8. $T_{\text{symbol}}^{c_i, {m_i}} = \frac{2^{m_i}}{BW_{c_i}}$ is the symbol time, and $BW_{c_i}$ is the bandwidth of CH $c_i$.

Since the LoRa network operates in the unlicensed spectrum, EDs should comply with the maximum restrictions of duty cycle $\delta $.
Due to the duty cycle, each transmission is divided into an active period, wherein packets are transmitted, as well as an inactive period, during which no packet transmission occurs. According to \cite{toro2021modeling}, the probability that the interfering ED $j$ is operating in the active period at a given time $T_j^{c_j,m_j}$ is defined as ${\delta _j} = 1 - 100(1 - \delta )\lambda T_j^{c_j,{m_j}}$.

By fully considering the effects of the duty cycle, the probability that interfering ED $j$ transmits in the interference time interval $T_{ij}$ can be given by:
\begin{equation}
\begin{aligned}
{h_j} = 1 - {e^{ - \lambda {T_{ij}}{\delta _j}}}
\end{aligned}
\end{equation}

\subsection{Capture Effect and Quasi-Orthogonality between SFs}

The signal transmitted by ED $i$ experiences channel fading and path loss before arriving at GW $k$, and the received signal strength is defined as:
\begin{equation}
\begin{aligned}
p_{i,k}^{\text{rx}} = {p_i} \cdot g_{i,k} \cdot a({d_{i,k}})
\end{aligned}
\end{equation}
where ${p_i}$ denotes the TP of ED $i$ and $g_{i, k}$ denotes the Rayleigh fading channel between GW $k$ and ED $i$. In addition, $a({d_{i,k}})$ denotes the path loss attenuation function, following the Friis transmission equation \cite{gao2019towards}, which can be expressed as:
\begin{equation}
\begin{aligned}
a({d_{i,k}}) = {\left (\frac{\omega }{{4\pi f{d_{i,k}}}}\right )^\tau }
\end{aligned}
\end{equation}
where $\omega $ and $f$ represent the light velocity the carrier frequency, respectively. $d_{i, k}$ is the distance from ED $i$ to GW $k$ and $\tau$ denotes the path loss exponent.

Owing to the capture effect and the quasi-orthogonal nature of SFs, a packet from the ED can be decoded if the received SIR is above a relative threshold, even in the presence of overlapping transmissions.
The thresholds for various SF pairs are given by the following SIR matrix \cite{mahmood2018scalability}:
\begin{equation}
\begin{aligned}
&\Delta_{[\text{dB}]} =
\begin{array}{*{20}{c}}
{SF_7}\\
{SF_8}\\
{SF_9}\\
{SF_{10}}\\
{SF_{11}}\\
{SF_{12}}
\end{array}\mathop {\left[ {\begin{array}{*{20}{c}}
		1&{ - 8}&{ - 9}&{ - 9}&{ - 9}&{ - 9}\\
		{ - 11}&1&{ - 11}&{ - 12}&{ - 13}&{ - 13}\\
		{ - 15}&{ - 13}&1&{ - 13}&{ - 14}&{ - 15}\\
		{ - 19}&{ - 18}&{ - 17}&1&{ - 17}&{ - 18}\\
		{ - 22}&{ - 22}&{ - 21}&{ - 20}&1&{ - 20}\\
		{ - 25}&{ - 25}&{ - 25}&{ - 24}&{ - 23}&1
		\end{array}} \right]}\limits^{\begin{array}{*{20}{c}}
	{SF_7}&{SF_8}&{SF_9}&{SF_{10}}&{SF_{11}}&{SF_{12}}
	\end{array}}\label{SIR matrix}
\end{aligned}
\end{equation}
The rows denote the SF selected by the target ED, and the columns denote the SF selected by the interfering ED.
The SIR of ED $i$ above the relevant threshold can be defined as follows:
\begin{equation}
\begin{aligned}
S_{(i,j),k} &= \dfrac{p_{i,k}^{\text{rx}}}{p_{j,k}^{\text{rx}}} = \frac{{p_i} \cdot g_{i,k} \cdot a({d_{i,k}}) }{{p_j} \cdot g_{j,k} \cdot a({x_{j,k}}) } \ge \eta^{\text{SIR}}_{m_i, m_j}
\end{aligned}\label{SIR}
\end{equation}
where $\eta^{\text{SIR}}_{m_i, m_j}$ denotes the SIR threshold that a packet is transmitted at $SF_{m_i}$ and the colliding packet is transmitted at $SF_{m_j}$.

\subsection{Packet Delivery Rate in Single-GW LoRa Networks}

According to \cite{gao2019towards}, GW $k$ can successfully decode an uplink packet transmitted from ED $i$ if the following two conditions are satisfied:

1) The received signal strength $p_{i,k}^{\text{rx}}$ exceeds the receiver sensitivity $\eta^{\text{sen}}_i$. The receiver sensitivity is determined according to Semtech specifications \cite{semtech2019sx1272}. The sensitivity thresholds for various SFs are defined in Table \ref{threshold_sen}.

2) The received SIR exceeds a certain threshold $\eta^{\text{SIR}}_{m_i, m_j}$ in the presence of overlapping transmissions.

In summary, the PDR of ED $i$ at GW $k$ is defined by Eq. (\ref{pdr_i_k}).

\begin{strip}\vspace{-5mm}
	\begin{equation}
	\begin{aligned}
	\begin{split}
	{D_{i,k}(\textbf{\textit{c}},\textbf{\textit{m}},\textbf{\textit{p}})} &= P\left\{p_{i,k}^{\text{rx}} \ge \eta^{\text{sen}}_{i} \right\} \times 
	\prod\limits_{c_i=c_j,\atop j\neq i}^{\cal N} X_j^{c_j,m_j}{\left[ h_j P\left\{ S_{(i,j),k} \ge \eta^{\text{SIR}}_{m_i, m_j} \right\} + \left( {1 - h_j} \right) \right]} \\
	& = P\left\{{p_i} \cdot g_{i,k} \cdot a({x_{i,k}}) \ge \eta^{\text{sen}}_{i} \right\} \times 
	\prod\limits_{c_i=c_j,\atop j\neq i}^{\cal N} X_j^{c_j,m_j} {\left[ h_j P\left\{ \frac{{p_i} \cdot g_{i,k} \cdot a({d_{i,k}}) }{{p_j} \cdot g_{j,k} \cdot a({x_{j,k}}) } \ge \eta^{\text{SIR}}_{m_i, m_j} \right\} + \left( {1 - h_j} \right) \right]} \\
	& = P\left\{ g_{i,k} \ge \dfrac{\eta^{\text{sen}}_{i}}{{p_i} \cdot a({d_{i,k}})}  \right\} \times 
	\prod\limits_{c_i=c_j,\atop j\neq i}^{\cal N} X_j^{c_j,m_j}{h_j}  {\left[ {P\left\{ g_{i,k} \ge \dfrac{\eta^{\text{SIR}}_{m_i, m_j}\cdot {p_j} \cdot g_{j,k} \cdot a({x_{j,k}})}{{p_i} \cdot a({d_{i,k}})} \right\} + \left( {1 - {h_j}} \right)} \right]} \\
	\end{split}
	\end{aligned}\label{pdr_i_k}
	\end{equation}\vspace{-0.5cm}
\end{strip}

For simplicity, we define $I_{(i,j),k}={p_j} \cdot g_{j,k} \cdot a({x_{j,k}})$ as the interference of ED $j$ with ED $i$.
Based on the Rayleigh fading assumption (i.e., $g\sim\exp(1)$), Eq. (\ref{pdr_i_k}) can be rewritten as:
\begin{equation}
\begin{aligned}
\begin{split}
&{D_{i,k}(\textbf{\textit{c}},\textbf{\textit{m}},\textbf{\textit{p}})} = 
\exp \left (- \dfrac{\eta^{\text{sen}}_{i}}{{p_i} \cdot a({d_{i,k}})}  \right ) \times \\
&\prod\limits_{c_i=c_j,\atop j\neq i}^{\cal N} X_j^{c_j,m_j} {\left[ {{h_j}  \exp \left ( - \dfrac{\eta^{\text{SIR}}_{m_i, m_j} I_{(i,j),k}}{{p_i} \cdot a({d_{i,k}})} \right ) + \left( {1 - {h_j}} \right)} \right]} \\
\end{split}
\end{aligned}
\end{equation}
where  
$\textbf{\textit{c}}=\{c_i\}_{i \in {\cal N}},\textbf{\textit{m}}=\{m_i\}_{i \in {\cal N}}$ and $\textbf{\textit{p}}=\{p_i\}_{i \in {\cal N}}$ represent the CH, SF and TP allocation for the EDs, respectively.
$X_j^{c_j,m_j}$ is a binary value. Here, $X_j^{c_j,m_j}=1$ means that SF $m_j$ and CH $c_j$ are allocated to ED $j$, and $X_j^{c_j,m_j}=0$ otherwise.

\begin{table}[h]
	\centering
	\caption{Sensitivity thresholds and distance ranges}
	\label{threshold_sen}
	\setlength{\tabcolsep}{1.75mm}{
		\begin{tabular}{ccccccc}
			\hline  
			& & & & & &\\[-5pt]
			\textbf{Spreading Factors}&7&8&9&10&11&12 \\
			\hline
			& & & & & &\\[-5pt]
			\textbf{Sensitivity (dBm)}&-123&-126&-129&-132&-134.5&-137\\
			\hline
			& & & & & &\\[-5pt]
			\textbf{Distance ranges (km)}&(0, 2]&(2,4]&(4,6]&(6,8]&(8,10]&(10,12]\\
			\hline
	\end{tabular}}\vspace{-0.5cm}
\end{table}  

\subsection{Packet Delivery Rate in Multi-GW LoRa Networks}
In LoRa networks, the broadcast packets from each ED can be decoded by any GW within the communication range. Then, the GW relays all the received packets to the NS, which discards duplicate packets. Therefore, a packet is successfully transmitted from an ED if it is received by at least one GW. Hence, the PDR of an ED in a multi-GW LoRa network is calculated as follows:
\begin{equation}
\begin{aligned}
{D_i}(\textbf{\textit{c}},\textbf{\textit{m}},\textbf{\textit{p}}) = 1 - \prod\limits_{k \in {\cal K}} {\left[ {1 - {D_{i,k}(\textbf{\textit{c}},\textbf{\textit{m}},\textbf{\textit{p}})}} \right]} 
\end{aligned}
\end{equation}
where $\prod\limits_{k \in {\cal K}} {\left[ {1 - {D_{i,k}(\textbf{\textit{c}},\textbf{\textit{m}},\textbf{\textit{p}})}} \right]}$ represents the probability that the packet from ED $i$ fails to be received by any GW.

\subsection{Energy Efficiency Model}
The EE of an ED can be defined as the number of transmitted information bits within an energy consumption unit. Therefore, the EE of ED $i$ is given by:
\begin{equation}
\begin{aligned}
EE_i(\textbf{\textit{c}},\textbf{\textit{m}},\textbf{\textit{p}}) = \frac{L}{{E_i^{\text{su}}}(\textbf{\textit{c}},\textbf{\textit{m}},\textbf{\textit{p}})} = \frac{L}{e_{p_i} \cdot T_i^{c_i,m_i} \cdot \frac{1}{D_i(\textbf{\textit{c}},\textbf{\textit{m}},\textbf{\textit{p}})}}
\end{aligned}
\end{equation}
where $E_i^{\text{su}}(\textbf{\textit{c}},\textbf{\textit{m}},\textbf{\textit{p}})$ represents the energy consumption required for a transmitting packet from ED $i$ to be successfully received by a GW. $e_{p_i}$ is the energy consumed during the transmission process in a time unit with TP $p_i$, and
$ \frac{e_{p_i} \cdot T_i^{c_i,m_i}}{D_i(\textbf{\textit{c}},\textbf{\textit{m}},\textbf{\textit{p}})}$ is the energy consumption due to interference. The interference may cause multiple retransmissions of a packet, subsequently leading to additional energy consumption.

Therefore, the system EE for uplink transmission in a LoRa network is defined as follows:
\begin{equation}
\begin{aligned}
EE(\textbf{\textit{c}},\textbf{\textit{m}},\textbf{\textit{p}}) = \sum\limits_{i \in {\cal N}} {E{E_i}(\textbf{\textit{c}},\textbf{\textit{m}},\textbf{\textit{p}})}
\end{aligned}
\end{equation}

\section{Problem Formulation}
We aim to maximize the system EE of uplink transmissions within multi-GW LoRa networks by jointly optimizing the CH assignment strategy $\textbf{\textit{c}}$, the SF assignment strategy $\textbf{\textit{m}}$ and the TP allocation strategy $\textbf{\textit{p}}$ for each ED. Formally, this allocation problem is defined as:
\begin{equation}
\begin{aligned}
{\textbf{P1}}:& \mathop {\max }\limits_{\textbf{\textit{c}}, \textbf{\textit{m}}, \textbf{\textit{p}}} EE(\textbf{\textit{c}},\textbf{\textit{m}},\textbf{\textit{p}}) \label{cmp}\\
s.t.\ &(\ref{cmp}{a})\ {p_{\min }} \le {p_i} \le {p_{\max }}, \forall i \in {\cal N} \\
&(\ref{cmp}{b})\  X_i^{c_i,m_i} \in \{0,1\}, \forall i \in {\cal N}, c_i \in {\cal C}, m_i \in {\cal M} \\
&(\ref{cmp}{c})\ \sum_{c_i \in {\cal C}} \sum_{m_i \in {\cal M}} X_i^{c_i,m_i} \le 1, \forall i \in {\cal N} \\
&(\ref{cmp}{d})\  \sum_{i \in {\cal N}_{c_i}} X_i^{c_i,m_i} \le \Lambda _{\max }, \forall c_i \in {\cal C}, m_i \in {\cal M} \\
&(\ref{cmp}{e})\  D_i(\textbf{\textit{c}},\textbf{\textit{m}},\textbf{\textit{p}}) \ge D_{\text{th}}, \forall i \in {\cal N} \\
\end{aligned}
\end{equation}
where constraint $(\ref{cmp}{a})$ represents the lower and upper bounds for the available TP. Constraint $(\ref{cmp}{b})$ defines the binary assignment variables $X_i^{c_i,m_i}$ for the CHs and SFs. Constraint $(\ref{cmp}{c})$ ensures that each ED is allowed to access at most one CH and select one SF.
Constraint $(\ref{cmp}{d})$ represents the upper limit on the number of EDs that can be assigned to each CH, and ${\cal N}_{c_i}$ is the group of EDs assigned to the same CH as ED $i$. Constraint $(\ref{cmp}{d})$ aims to reduce the implementation interference and complexity of each CH.
Constraint $(\ref{cmp}{e})$ specifies the PDR threshold $D_{\text{th}}$ to guarantee reliable
transmission.

\begin{figure}[tb]
	\centering
	\includegraphics[width=90mm]{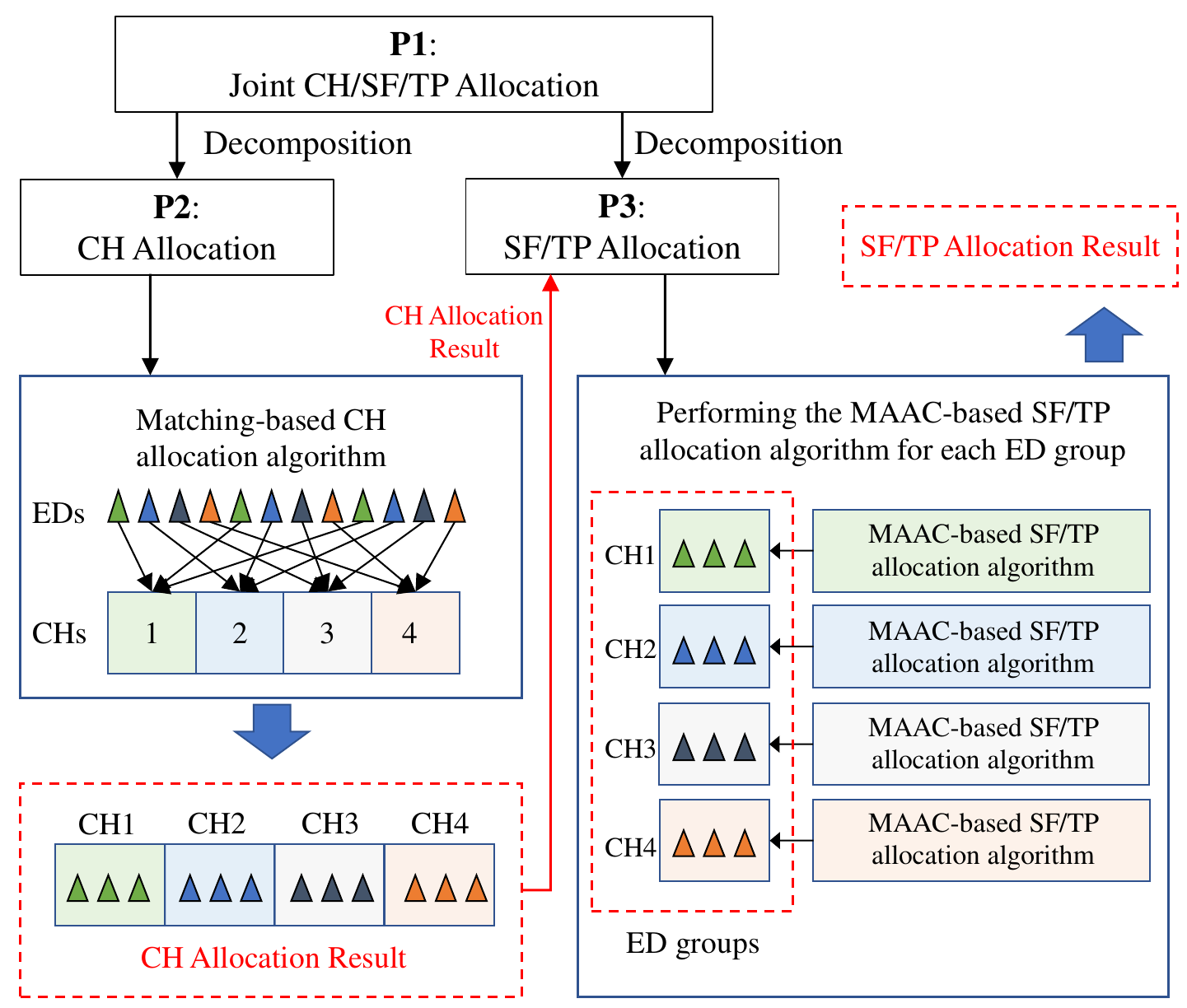}
	\caption{The workflow of the MMALoRa framework.
	}\label{MMALoRa_framework}
\end{figure}

Due to the coupling of the SF/TP/CH assignments and the nonconvex optimization objective, the problem described is NP-hard, indicating that resolving this problem within polynomial time is challenging.
Therefore, a two-stage optimization framework, MMALoRa, is proposed to derive the solution to problem $\textbf{P1}$ in two stages: CH assignment and SF/TP assignment. The workflow of MMALoRa is illustrated in Fig. \ref{MMALoRa_framework}. In the initial stage, a matching-based algorithm is employed to match the CH for each ED. Given the CH assignment, all EDs can be segmented into separate groups according to the EDs assigned to the same CH. In the final stage, an MAAC-based algorithm is employed to allocate SF/TP values for each ED group in parallel.

\section{Matching-Based CH Assignment Algorithm}
In this section, we present a matching-based CH assignment algorithm, with the EDs and CHs regarded as two distinct sets of players to be matched to optimize the EE.

\subsection{CH Assignment Optimization}
It is assumed that each ED is assigned the maximum TP and  a fixed SF. The SF allocation is determined based on a predefined relationship between the distance from the ED to the GW and the corresponding SF, as specified in Table \ref{threshold_sen}.
Given that the PDR of an ED is heavily influenced by its SF/TP allocation, adjusting the ED's PDR solely through CH assignment is likely to be ineffective. Consequently, we do not account for the PDR constraints of the ED during the CH assignment process.
Thus, problem $\textbf{P1}$ can be reformulated as follows:
\begin{equation}
\begin{aligned}
{ \textbf{P2}}:& \mathop {\max }\limits_{\textbf{\textit{c}}} EE(\textbf{\textit{c}}) \label{c} \\
s.t.\ &(\ref{c}{a})\  X_i^{c_i,m_i} \in \{0,1\}, \forall i \in {\cal N}, c_i \in {\cal C}, m_i \in {\cal M}  \\
&(\ref{c}{b})\ \sum_{c_i \in {\cal C}} \sum_{m_i \in {\cal M}} X_i^{c_i,m_i} \le 1, \forall i \in {\cal N} \\
&(\ref{c}{c})\  \sum_{i \in {\cal N}_{c_i}} X_i^{c_i,m_i} \le \Lambda _{\max }, \forall c_i \in {\cal C}, m_i \in {\cal M} \\
\end{aligned}
\end{equation}

Problem $\textbf{P2}$ is classified as a many-to-one matching problem since at most one CH can be matched with an ED, while multiple EDs can be matched with the same CH.
Moreover, due to the interference term in Eq. (\ref{SIR}), the EE of any ED over its occupied CH is related to other EDs sharing this CH. 
Therefore, for each ED, the matched CH and the other EDs matched with the same CH must both be considered. Similarly, for each CH, not only the EDs matched with this CH but also the interrelationships within the group of EDs matched with this CH, which are influenced by cochannel interference (i.e., inter-SF and co-SF interference in the same CH, where co-SF is the interference generated between EDs that select the same SF on the same CH) must be considered. 
Thus, problem $\textbf{P2}$ also includes externalities.

In traditional matching, each set of players establishes a fixed preference ordering based on their preferences for the other set of players \cite{gale1962college}. Next, players from different sets are matched with each other according to their preference ordering. However, the presence of externalities leads to changes in the preference ordering of players as matching progresses, which considerably complicates the matching process. To address these concerns, we employ the approach of swap matching, as described in \cite{bodine2011peer}, which captures the externalities using utility functions instead of traditional preference ordering.

To more effectively describe the swap matching approach used in the many-to-one matching problem with externalities, we initially define the foundational concepts of matching theory:

\textit{1) Notation for many-to-one matchings:}
We define two disjoint sets, the ED set $\cal N$ and the CH set $\cal C$. For the CH set, there exists a positive integer quota $\Lambda _{\max}$, which denotes the maximum limit on the number of EDs using the same CH and is defined in (\ref{c}{$c$}).
A matching $\psi$ describes the allocation of EDs to CHs such that each ED is matched with only one CH, while each CH is matched with multiple EDs. More formally, this can be represented as follows:

\textit{\textbf{Definition 1:} A many-to-one matching is a subset $\psi \subseteq \cal C \times \cal N$ such that $|\psi(i)| = 1$ and $|\psi(c)| \leq \Lambda _{\max}$, where $\psi(i) = \{c \in \cal C : $ $ (i, c)\in \psi \} $ and $\psi(c) = \{i \in \cal N :$ $(i, c)\in \psi \} $.}

\textit{2) Utility function:} 
Each CH prefers to be matched with the group of EDs with the largest utility, whereas each ED prefers to be matched with the CH that offers the largest utility.

For a given matching $\psi$, the utility of ED $i$ under $\psi$ can be defined as follows:
\begin{equation}
\begin{aligned}
{U_i}(\psi ) = E{E_i}(\psi )
\end{aligned}\label{utility_ED}
\end{equation}

The utility of CH $c$ under the matching $\psi$ is given by:
\begin{equation}
\begin{aligned}
{U_c}(\psi ) = \sum\limits_{i \in {{\cal N}_c}} {E{E_i}}(\psi )
\end{aligned}\label{utility_CH}
\end{equation}
where ${\cal N}_c$ is the ED group assigned to CH $c$.

\textit{3) Two-sided exchange stable (\textbf{2ES}):} Externalities make designing matching mechanisms more challenging, as stable matching cannot be guaranteed straightforwardly.
Therefore, \textbf{2ES} is introduced to describe the effect of externalities on the matching outcomes \cite{bodine2011peer}.
To better describe \textbf{2ES}, the principle of swap matching is initially introduced, defined as follows:

\textit{\textbf{Definition 2}: Swap matching behaviour $\psi _i^{i'} = \psi \backslash \{ (i,c),(i',c')\}  \cup \{ (i',c),(i,c')\}$ is defined by $\psi _i^{i'}(i)=c'$ and $\psi _i^{i'}(i')=c$.}

Swap matching is accomplished through the execution of swap operations; thus, the notion of swap-blocking pairs is introduced.

\textit{\textbf{Definition 3}: Given a pair $(i, i')$ that is matched according to $\psi$, if there exist $\psi (i)=c$ and $\psi (i')=c'$ such that\\
	1) \ \ $\forall q \in \{ i,i',\psi (i),\psi (i')\} ,{U_q}(\psi _i^{i'}) \ge {U_q}(\psi)$ \\
	2) \ \ $\exists q \in \{ i,i',\psi (i),\psi (i')\} ,{U_q}(\psi _i^{i'}) > {U_q}(\psi)$\\
	then $(i,i')$ is a swap-blocking pair. }

Under condition \textit{1)}, the utility of any involved player $q$ cannot decrease after performing the swap operation between $(i, i')$.
Under condition \textit{2)}, the utility of at least one player $q$ can improve after performing the swap operation between $(i, i')$.
Then, \textbf{2ES} matching can be achieved after multiple swap matching operations, which is described in the following definition:

\textit{\textbf{Definition 4}: If there are no swap-blocking pairs, the matching $\psi$ is \textbf{2ES}.}

\subsection{Proposed Matching-Based Algorithm}
The matching-based CH assignment algorithm consists of an initialization phase and a swap matching phase, which is summarized in Algorithm \ref{MCA}. First, we give an initial matching $\psi_0$, in which EDs and CHs are randomly matched with each other. Next, each ED constantly searches for swap-blocking pairs to execute the swap operation until no swap-blocking pairs remain. Finally, the final stable matching state is returned.

\begin{algorithm}
	\caption{Matching-based CH assignment algorithm }
	\label{matching}
	\begin{algorithmic}[1]
		\STATE \textbf{Step 1: Initialization}
		
		\STATE Initial matching $\psi = \psi_0$: EDs and CHs are randomly matched with each other under constrains of $|\psi(i)| = 1$ and $|\psi(c)| \leq \Lambda _{\max}$;
		
		\STATE \textbf{Step 2: Swap matching}
		\REPEAT               
		\FOR{$\forall i \in \mathcal{N}$}
		
		\FOR{$\forall i' \in \mathcal{N}_{-i}$, with $\psi (i)=c$ and $\psi (i')=c'$}
		\IF{$(i,i')$ is a swap-blocking pair}
		\STATE Execute the swap operation;
		\STATE Update the current matching: $\psi=\psi _i^{i'}$;
		\ELSE 
		\STATE Keep the current matching;
		\ENDIF
		\ENDFOR
		
		\ENDFOR           
		\UNTIL{no swap-blocking pair exists.}
		\STATE \textbf{Return the final matching} $\psi^* $
	\end{algorithmic}\label{MCA}
\end{algorithm}

\textit{\textbf{Theorem 1}: Through a limited number of swap operations, the proposed matching-based CH assignment algorithm is capable of converging to a \textbf{2ES} matching.}

\textit{Proof}: After each swap operation between ED $i$ and ED $i'$ with $i\in \psi(c)$, $i'\in \psi(c')$, the EDs are matched with CH $c$ and $c'$, respectively, which changes the matching from $\psi$ to $\psi_i^{i'}$. According to \textbf{\textit{Definition 3}}, the utility of any player among $\{i, i', c, c' \}$ does not decrease when a swap operation is performed and the utility of at least one player can be augmented.
Therefore, the EE of at least one player will improve after performing a swap operation. On the one hand, the finite number of EDs and CHs restricts the number of swap operations per iteration. On the other hand, the EE is constrained by an upper limit owing to the finite spectrum resources \cite{su2020energy}. As a result, the number of iterations is finite and the proposed matching-based algorithm can converge to a \textbf{2ES} matching within a limited number of swap operations.

\textit{\textbf{Theorem 2}: Assuming a total iteration count of $I$, the computational complexity of the matching-based algorithm is bounded by $O(I{\Lambda^2 _{\max }} C (C - 1))$.}

\textit{Proof}: For any CH $c$, the algorithm evaluates a maximum of ${\Lambda _{\max }}$ EDs and inspects $\Lambda _{\max } (C - 1)$ potential swap operations for each of these EDs.
Therefore, there exist ${\Lambda^2 _{\max }} C (C - 1)$ possible swap operations in one iteration. Consequently,  given the number of iterations $I$, the complexity of the matching-based algorithm is bounded by $O(I{\Lambda^2 _{\max }} C (C - 1))$.

\section{MAAC-Based SF/TP Allocation Algorithm}

In this section, we optimize the SF/TP assignments for EDs allocated to the same CH to maximize the EE. First, we formulate the SF/TP allocation problem. Subsequently, the aforementioned problem is reformulated as a Markov game, with the training and execution phases of the MAAC-based algorithm presented in detail.

\subsection{SF/TP Allocation Optimization}

Given the CH assignment, \textbf{P1} is defined as follows:
\begin{equation}
\begin{aligned}
{\textbf{P3}}:& \mathop {\max }\limits_{\textbf{\textit{m}},\textbf{\textit{p}}} EE(\textbf{\textit{m}},\textbf{\textit{p}}) = \mathop {\max }\limits_{\textbf{\textit{m}},\textbf{\textit{p}}} \sum_{c \in \cal C} EE_c(\textbf{\textit{m}},\textbf{\textit{p}}) \label{mp}\\
s.t.\  &(\ref{mp}{a})\ {p_{\min }} \le {p_i} \le {p_{\max }}, \forall i \in {\cal N} \\
&(\ref{mp}{b})\  X_i^{c_i,m_i} \in \{0,1\}, \forall i \in {\cal N}, c_i \in {\cal C}, m_i \in {\cal M} \\
&(\ref{mp}{c})\ \sum_{c_i \in {\cal C}} \sum_{m_i \in {\cal M}} X_i^{c_i,m_i} \le 1, \forall i \in {\cal N} \\
&(\ref{mp}{d})\ D_i(\textbf{\textit{m}},\textbf{\textit{p}}) \ge D_{\text{th}}, \forall i \in {\cal N} \\
\end{aligned}
\end{equation}

Note that obtaining the optimal solution for \textbf{P3} remains challenging due to the coupling of the SF and TP assignments and the nonconvex optimization objective. To address this problem, an MAAC-based algorithm is employed to determine the SF/TP allocation for the EDs to maximize the EE in each CH.

Specifically, we initially recast \textbf{P3} as a Markov game with multiple EDs, which is defined as a tuple $\langle {\cal N}, {\cal S}, {\cal T}, \{{\cal A}_i\}_{i\in {\cal N}}, \{{\cal R}_i\}_{i\in {\cal N}} \rangle$, where
$\cal S$ represents a set of states,
$\cal T$ denotes the state transition function, ${\cal R}_i$ is the reward function space of ED $i$ and ${\cal A}_i$ denotes the action space of ED $i$.
At time slot $t$, ED $i$ selects action $a_i^t \in {\cal A}_i$ based on local observation $o_i^t$, where $o_i^t$ is the proportion of the global state $s^t \in \cal S$. Then, ED $i$ receives a reward $r_i^{t+1}$, and the state transits from $s^{t}$ to $s^{t+1}$ with probability ${\cal T}(s^{t+1}|s^t,a_1^t, \cdots, a_N^t)$. Each ED $i$ attempts to learn the optimal policy $\pi_i$, which maximizes its cumulative discounted reward $\sum_{n=0}^{\infty} \mu^n r_i^{t+1+n}$, where $\mu$ denotes the discount rate.
Given that the MAAC-based algorithm is a model-free reinforcement learning approach, information on the state transition function is not needed. Consequently, we concentrate on the design of the three fundamental elements of the Markov game pertaining to \textbf{P3}: the state, action, and reward function.

\textbf{State}: The observation space for each ED includes three distinct components: the individual PDR, the individual EE, and the distances between the ED and each GW. Since the transmission performance of each ED cannot be determined before the initiation of time slot $t$, the ED relies on delayed transmission performance as a crucial environmental feature. At time slot $t$, the observation space of agent $i$ is given by:
\begin{equation}
\begin{aligned}
o_i^t=\{D_i^{t-1},EE_i^{t-1},\textit{\textbf{d}}_i\}
\end{aligned}\label{state}
\end{equation}
where $\textit{\textbf{d}}_i=\{d_{i,1}, d_{i,2}, \cdots, d_{i,K}\}$ is the set of distances from ED $i$ to each GW.
Without loss of generality, the global state $s^t$ is selected to be the set of observations of all EDs $\textbf{\textit{o}}^t=\{o_1^t, \cdots, o_N^t \}$.

\textbf{Action}: By receiving the current observation at time slot $t$, each agent determines the TP/SF allocation.
The TP of each ED is discretized into $J$ levels, and the power level set is given by:
\begin{equation}
\begin{aligned}
&{\cal P}=\\
&\left\{p_{\text{min}},p_{\text{min}}+\frac{p_{\text{max}}-p_{\text{min}}}{J-1},p_{\text{min}}+\frac{2(p_{\text{max}}-p_{\text{min}})}{J-1},\cdots,\ p_{\text{max}}\right\}
\end{aligned}\label{action}
\end{equation}

Then, the action of agent $i$ is defined as:
\begin{equation}
\begin{aligned}
a_i^t=\{m_i^t, p_i^t\}
\end{aligned}
\end{equation}
where $ p_i^t\in{\cal P}$, $ m_i^t \in {\cal M}$. Consequently, the action space dimension for each ED is $|\cal M|\times |\cal P|$.

\textbf{Reward}: Our proposed reward function is designed to optimize the nonconvex objective specified in problem \textbf{P3}.

The reward for ED $i$ is composed of two parts: (1) the total EE of all EDs that use the same CH $c$ as ED $i$, i.e., $EE_{c}$, and (2) the average $EE_{c}$ reduction attributed to interference from ED $i$, i.e., $\overline{EE}_{(c,-i)}$.
Therefore, at time slot $t$, given the action $a_i^t=\{m_i^t, p_i^t\}$ of agent $i$, the reward for ED $i$ can be defined as:
\begin{equation}
\begin{aligned}
&r_i^{t+1} = \gamma_{D_{\text{th}}} \cdot \left [\varrho EE_{c} + (1-\varrho){\overline{EE}_{({c},-i)}}\right ] \\
&=\gamma_{D_{\text{th}}} \cdot \left [ \varrho \sum\limits_{i \in {{\cal N}_{c}}} {E{E_i}} + (1-\varrho) (\dfrac{EE_{c}}{N} - \dfrac{EE_{({c},-i)}}{N-1})\right ]
\end{aligned}\label{reward} 
\end{equation}
where $\varrho$ represents a nonnegative weight that balances the trade-off between the system EE and the penalty imposed on ED $i$. $EE_{({c},-i)}$ denotes the EE of all EDs in CH $c$ without considering ED $i$, and $\gamma_{D_{\text{th}}}$ is a binary value used to ensure that the PDR constraint is satisfied, which can be given by: 
\begin{equation}
\gamma_{D_{\text{th}}}=\left\{
\begin{aligned}
1 & , & D_i \ge D_{\text{th}} \\
0 & , & D_i < D_{\text{th}}
\end{aligned}
\right.
\end{equation} 

\subsection{Proposed MAAC-Based Algorithm}

\begin{figure*}[htbp]
	\vspace{-0.3cm}
	\centering
	\includegraphics[width=180mm]{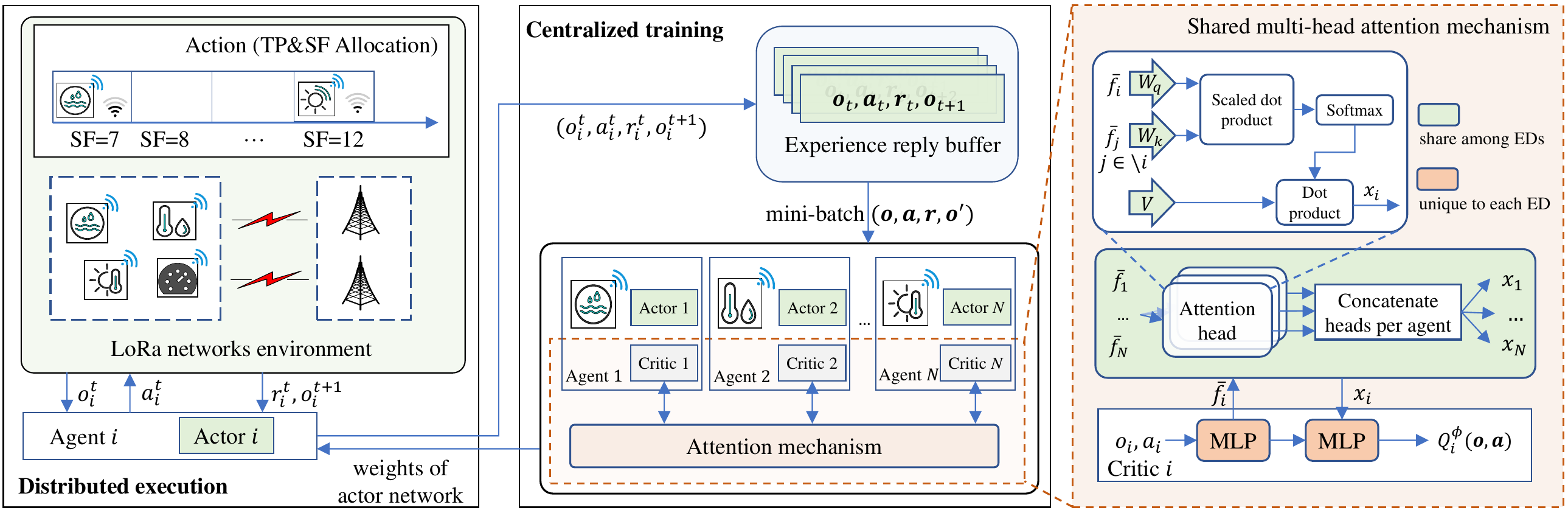}
	\caption{The MAAC-based algorithm architecture for SF/TP allocation.
	}\label{MMALoRa}
\end{figure*}

The MAAC-based algorithm adopts a soft actor-critic (SAC) framework and incorporates a multihead attention mechanism.
Fig. \ref{MMALoRa} shows the MAAC-based algorithm architecture, which consists of a centralized training module and a distributed execution module. Since the NS can obtain the ED information collected by all GWs, centralized training is performed on the NS. This architecture reduces the signalling overhead caused by information transmission among the EDs, thereby increasing the flexibility of the algorithm. A complete description of the MAAC-based algorithm is provided in Algorithm \ref{AMSTA}.

In the \textbf{distributed execution} module, each ED acts as an agent and interacts with the LoRa network in each time slot $t$. Agent $i$ $(\forall i \in {\cal N})$ selects action $a_i^t$ based on the current observation $o_i^t$. Then, agent $i$ receives a new observation $o_i^{t+1}$ as well as a corresponding reward $r_i^{t+1}$, and the transition tuple $(o_i^t, a_i^t, r_i^{t+1}, o_i^{t+1})$ is collected. After collecting the transition tuple of each agent, the transition tuples of all agents $(\textbf{\textit{o}}^t, \textbf{\textit{a}}^t, \textbf{\textit{r}}^{t+1}, \textbf{\textit{o}}^{t+1})$ are stored in the experience replay buffer $\mathcal{B}$, where $\textbf{\textit{a}}^t=(a_1^t, \ldots, a_N^t)$, $\textbf{\textit{o}}^t=(o_1^t, \ldots, o_N^t)$, $\textbf{\textit{o}}^{t+1}=(o_1^{t+1}, \ldots, o_N^{t+1})$ and $\textbf{\textit{r}}^{t+1}=(r_1^{t+1}, \ldots, r_N^{t+1})$.

In the \textbf{centralized training} module, if the length of the experience replay buffer $l_{\mathcal{B}}$ exceeds the batch size $B$, the mini-batch tuples $\textbf{\textit{B}}_{\text{mini}}$ are randomly sampled from the experience replay buffer $\mathcal{B}$ in every time slot $T_{\text{update}}$ to train the policy.
Because the transmission parameters assigned to the EDs are tightly coupled, the performance of an individual ED can be significantly influenced by the actions of other EDs.
To address this problem, a multihead attention mechanism is employed to learn the critic for each agent by selectively considering information from other agents. Specifically, each agent incorporates the observations and actions of other agents into its own action-value function $Q_i^\phi(\textit{\textbf{o}},\textit{\textbf{a}})$, which is given by:
\begin{equation}
\begin{aligned}
Q_i^\phi(\textbf{\textit{o}},\textbf{\textit{a}})=F_i({\bar{f}}_i,x_i)
\end{aligned}
\end{equation}
where $\phi$ denotes the weight parameter of the critic network, $F_i$ denotes a two-layer multilayer perceptron (MLP), and ${\bar{f}}_i=f_i(o_i,a_i)$ denotes a one-layer MLP embedding function. In addition, $x_i$ denotes the contribution of other agents and is defined as follows:
\begin{equation}
\begin{aligned}
x_i=\sum_{j\neq i}{\rho_jh(V{\bar{f}}_j)}
\end{aligned}
\end{equation}
where the shared matrix $V$ transforms the embedding function ${\bar{f}}_j$ into a ``value'' and $h$ denotes a leaky ReLU activation function. $\rho_j$ is the attention weight. A query-key system is used to compare the similarity between ${\bar{f}}_j$ and ${\bar{f}}_i$; then, the similarity value is passed to a softmax function. $\rho_j$ can be calculated as follows:
\begin{equation}
\begin{aligned}
\rho_j=\frac{\exp{({\bar{f}}_j^TW_k^TW_q{\bar{f}}_i)}}{\sum_{j=1}^{N}\exp{({\bar{f}}_j^TW_k^TW_q{\bar{f}}_i)}}
\end{aligned}
\end{equation}
where ${\bar{f}}_j$ is transformed into a ``key'' by $W_k$, and ${\bar{f}}_i$ is transformed into a ``query'' by $W_q$. Note that we use a multihead attention mechanism, where each head uses a separate parameter group $(V, W_k, W_q)$ to calculate the contributions of other agents to agent $i$. Due to the sharing of the separate parameter groups, all critic networks are updated to minimize a joint regression loss function, which is defined as follows:
\begin{equation}
\begin{aligned}
\mathcal{L}_Q\ (\phi)=\sum_{i=1}^N\mathbb{E}_{(\textbf{\textit{o}}, \textbf{\textit{a}}, \textbf{\textit{r}}, \textbf{\textit{o}}')\sim D}[\left(Q_i^\phi(\textbf{\textit{o}},\textbf{\textit{a}})-y_i\right)^2],
\end{aligned}\label{update_critic}
\end{equation}
where $y_i=r_i+\mu\mathbb{E}_{\textbf{\textit{a}}'\sim\pi_{\bar{\theta}}}[-\alpha\log{(\pi_{\bar{\theta_i}}(a_i'|o_i'))}+Q_i^{\bar{\phi}}(\textbf{\textit{o}}',\textbf{\textit{a}}')]$,
$\bar{\phi}$ and $\bar{\theta}$ are the weights of the target critic network and target actor network, respectively, $\alpha$ denotes the temperature parameter and is used to determine the balance between exploitation and exploration within the SAC framework.

The individual policies are updated with the gradient ascent algorithm, and the gradient can be defined as follows:
\begin{equation}
\begin{aligned}
\nabla_{\theta_i}J(\theta)=\mathbb{E}_{\textbf{\textit{o}}\sim\mathcal{B},\textbf{\textit{a}}\sim\pi}[\nabla_{\theta_i}\log{(\pi_{\theta_i}(a_i|o_i))}\varphi_i(a_i,o_i)]
\end{aligned}\label{update_actor}
\end{equation}
where $\varphi_i(a_i,o_i)=-\alpha\log{(\pi_{\theta_i}(a_i|o_i))+A_i(\textbf{\textit{o}},\textbf{\textit{a}})}$; the multiagent advantage function, $A_i(\textbf{\textit{o}},\textbf{\textit{a}})$, is used to indicate whether the action of agent $i$ could improve its expected return, which can be defined as:
\begin{equation}
\begin{aligned}
A_i(\textbf{\textit{o}},\textbf{\textit{a}})=Q_i^\phi(\textbf{\textit{o}},\textbf{\textit{a}})-b(\textbf{\textit{o}},\textbf{\textit{a}}_{-i})
\end{aligned}
\end{equation}
where $\textbf{\textit{a}}_{-i}$ represents the actions of all agents except agent $i$. $b(\textbf{\textit{o}},\textbf{\textit{a}}_{-i})$ represents the baseline of the multiagent strategy, which can be defined as $b(\textbf{\textit{o}},\textbf{\textit{a}}_{-i}) = \sum_{a'_i \in \mathcal{A}_i} \pi_{\theta_i} (a'_i|o_i)Q_i^{\phi}(\textbf{\textit{o}},(a'_i,\textbf{\textit{a}}_{-i}))$.

\begin{algorithm}[tb]
	\caption{MAAC-based SF/TP allocation algorithm}
	\begin{algorithmic}
		\STATE \textbf{Initialize:\\}
		\STATE The experience replay buffer $\mathcal{B}$ \\
		\STATE The weights of critic network $\bar{\phi} \gets \phi$\\
		\STATE The weights of actor network $\bar{\theta} \gets \theta$   \\
		\FOR {$\text{episode}=1, 2, \cdots, N_{\text{episode}}$} 
		\STATE Reset the operating environment \\
		\STATE Each agent $i$ gets the initial observation state $o_i^t$ 
		\FOR {$t=1,2,\cdots, T_{\text{train}}$} 
		\STATE Each agent $i$ takes the action $a_i^t\sim\pi_{\theta_i}(o_i^t)$ \\
		\STATE Each agent $i$ calculates $r_i^{t}$ according to Eq. (\ref{reward})\\
		\STATE  Each agent $i$ observes $o_i^{t+1}$\\
		\STATE Store $(\textbf{\textit{o}}^t, \textbf{\textit{a}}^t, \textbf{\textit{r}}^{t}, \textbf{\textit{o}}^{t+1})$ for all agents in $\mathcal{B}$
		\ENDFOR
		\IF {$l_{\mathcal{B}}>B$ and $T_{\text{update}} \% t = 0$} 
		\STATE {Randomly sample mini-batch $\textbf{\textit{B}}_{\text{mini}}$ from $\mathcal{B}$\\
			\STATE Minimizing the joint regression loss function to update the critic networks according to Eq. (\ref{update_critic})\\
			\STATE Performing gradient ascent to update the actor networks according to Eq. (\ref{update_actor}) \\
			\STATE Update the target actor networks: $\bar{\theta} \gets \zeta\theta + (1-\zeta)\bar{\theta}$ \\
			\STATE Update the target critic networks:$\bar{\phi} \gets \zeta\phi + (1-\zeta)\bar{\phi}$} 
		\ENDIF
		\ENDFOR
	\end{algorithmic}\label{AMSTA}	
\end{algorithm}

\section{Numerical Results}
In this section, the impact of various parameter settings on the accuracy of the proposed analytical model is first analyzed.
Subsequently, the performance of the MMALoRa algorithm and the effects of various parameter settings on the LoRa network's performance are evaluated.

\subsection{Simulation Setup}

\begin{figure}[tb]
	\centering
	\includegraphics[width=80mm]{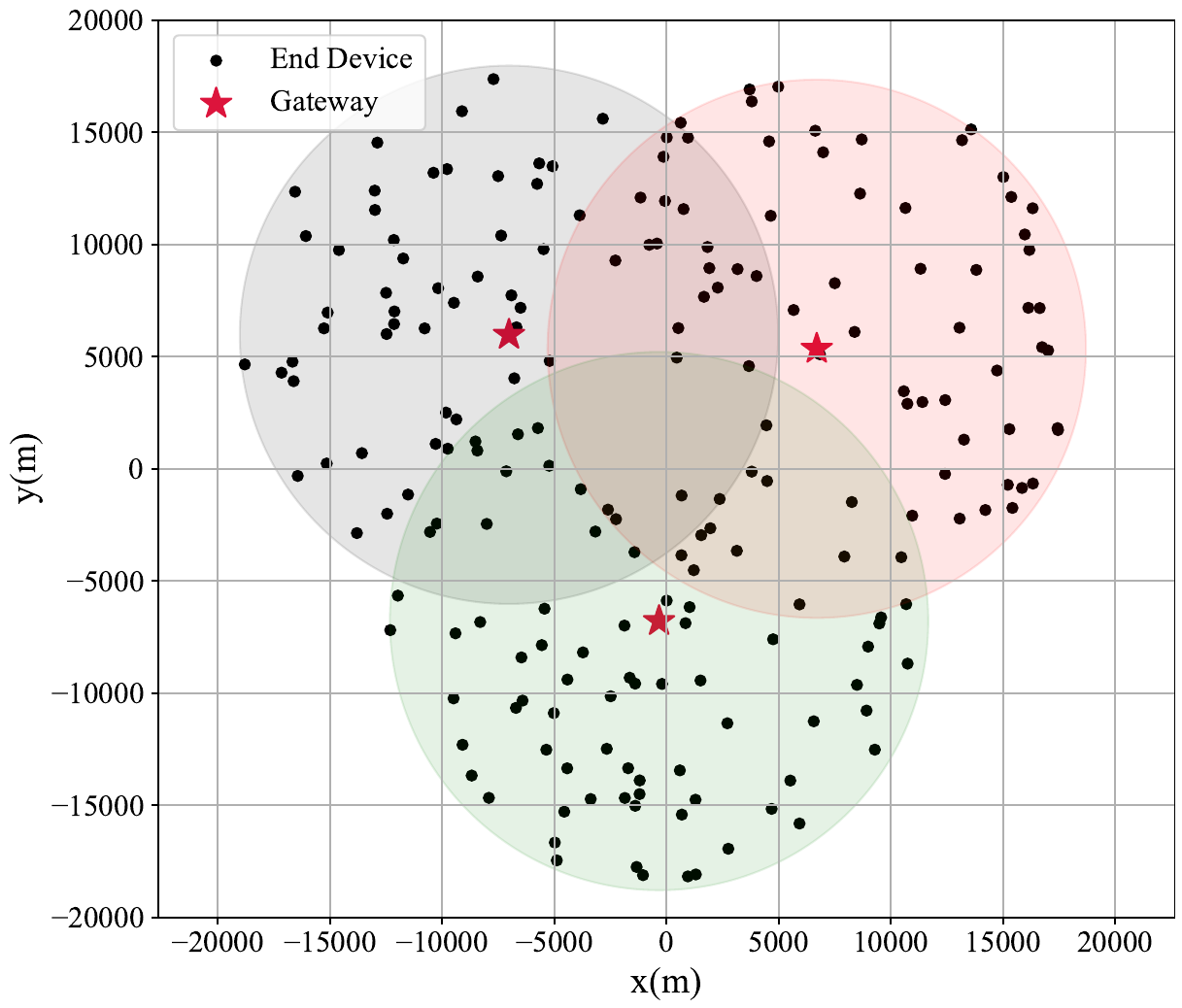}
	\caption{Locations of the GWs and EDs.
	}\label{network_modle}
\end{figure}

In the simulation setup, as outlined by the GW locations described in \cite{toro2021modeling}, the GWs are not deployed too close to each other while allowing for overlapping coverage areas.
Therefore, multiple GWs are randomly distributed within an area of 20 km $\times$ 20 km, with a minimum separation of 12 km between each GW. Moreover, each GW serves as the center of a circular cell with a radius of $R=12$ km. All EDs are uniformly distributed within these circular cells to ensure that all EDs can reach at least one GW. The locations of the GWs and EDs are shown in Fig. \ref{network_modle}. All the GWs and EDs are configured to operate at the 868 MHz ISM band channel frequency, with a bandwidth of 125 kHz. The duty cycle is 1\%, and the payload size is 20 bytes. The remaining simulation parameters are summarized in Table \ref{parameter}.

\begin{table}[h]
	\captionof{table}{Simulation parameters}
	\begin{tabular}{p{5cm}|p{3cm}}
		\hline
		\textbf{Parameters} & \textbf{Values} \\
		\hline
		Path loss exponent ($\tau$) & 2.7 \\
		\hline
		Average sending rate ($\lambda$) & 0.001 $s^{-1}$ \\
		\hline
		Preamble length ($n_{\text{pr}}$) & 8 bytes \\
		\hline
		Coding rate & 4/5 \\
		\hline
		Minimum transmission power limit ($p_{\text{min}}$) & 2 dBm \\
		\hline
		Maximum transmission power limit ($p_{\text{max}}$) & 20 dBm \\
		\hline
		Reward weights ($\varrho$) & $1/N$ \\
		\hline
	\end{tabular}\label{parameter}
\end{table}

\begin{table}
	\captionof{table}{The training parameters of the MMALoRa algorithm}
	\begin{tabular}{p{5cm}|p{3cm}}
		\hline
		\textbf{Parameters} & \textbf{Values} \\
		\hline
		Length of an episode ($T_{\text{train}}$) & 30 \\
		\hline
		Size of experience replay buffer ($|\mathcal{B}|$) & $10^5$ \\
		\hline
		Minibatch size ($\textbf{\textit{B}}_{\text{mini}}$) & 1024 \\
		\hline
		Learning rate  & 0.001 \\
		\hline
		Discount rate ($\mu$) & 0.99 \\
		\hline
		Target network update rate ($\zeta$) & 0.001 \\
		\hline
		Number of attention heads & 2 \\
		\hline
	\end{tabular}\label{train_parameter}
\end{table}

\subsection{Model Accuracy Analysis}
In this subsection, we verify the accuracy of the analytical model in three scenarios, which vary in the number of EDs, the number of GWs, and the transmission parameter configurations.
Similar to \cite{toro2021modeling},  it is assumed that all EDs utilize the same CH and the highest TP.
Since the EE of an ED is related mainly to the PDR, we compare the PDR derived by the proposed analytical model with that generated by the open-source LoRasim simulator \cite{bor2016lora} to evaluate the accuracy of our analytical model.
The PDR within the LoRasim simulator is defined as the ratio of the number of successfully received packets to the total transmitted packets by an ED. 

\begin{figure*}[tb]  
	\centering
	\subfigure[]{
		\label{MAE_ED}
		\begin{minipage}[t]{0.33\linewidth}
			\centering
			\includegraphics[width=60mm]{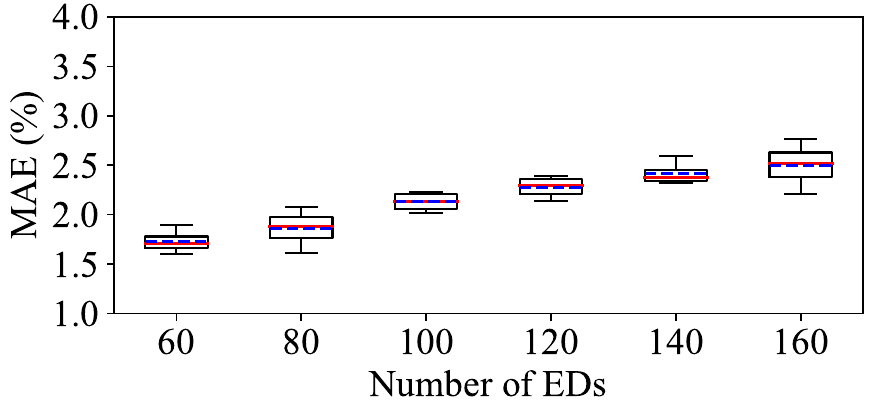}
		\end{minipage}%
	}%
	\subfigure[]{
		\label{MAE_GW}
		\begin{minipage}[t]{0.33\linewidth}
			\centering
			\includegraphics[width=60mm]{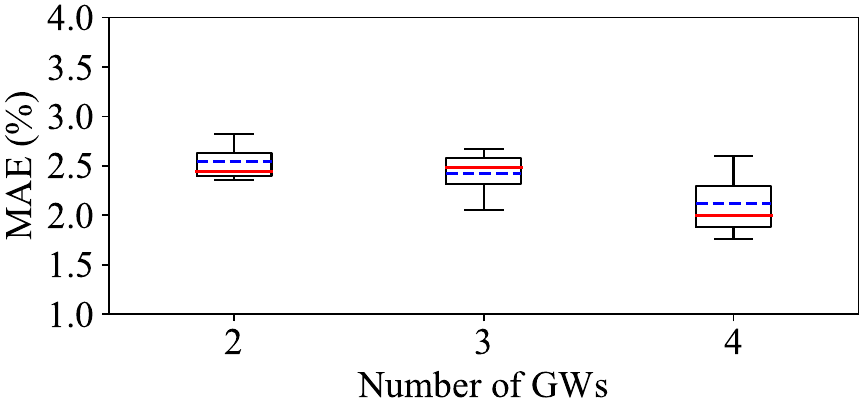}
		\end{minipage}%
	}%
	\subfigure[]{
		\label{MAE_algorithm}
		\begin{minipage}[t]{0.33\linewidth}
			\centering
			\includegraphics[width=58mm]{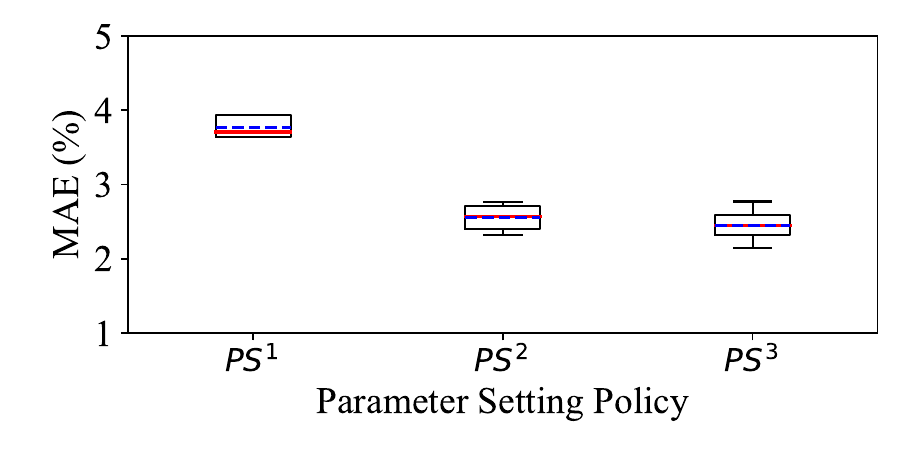}
		\end{minipage}%
	}%
	\centering
	\caption{MAE between the proposed analytical model and simulations for various (a) numbers of EDs, (b) numbers of GWs and (c) transmission parameter settings.}
	\vspace{-0.5cm}
	\label{MAE}
\end{figure*}

We use the mean absolute error (MAE) to evaluate the accuracy of the proposed analytical model:
\begin{equation}
\begin{aligned}
\text{MAE} = \dfrac{1}{|\cal N|}\sum\limits_{i \in {\cal N}} |D_i - \hat{D}_i|
\end{aligned}
\end{equation}
where $\cal N$ is the ED set and $D_i$ and $\hat{D}_i$ are the PDRs of ED $i$ calculated by the proposed analytical model and the LoRasim simulator, respectively. Fig. \ref{MAE} shows the MAEs with a confidence interval of 95\%, where the blue dashed line is the mean and the red solid line is the median. 

First, the accuracy of the analytical model is evaluated for networks with various numbers of EDs ranging from 60 to 160, and the number of GWs is fixed at 3. All the EDs are assigned the largest SF.
As shown in Fig. \ref{MAE_ED}, the MAE is less than 3\% and increases slightly with the increasing number of EDs. This outcome occurs owing to the inherent stochastic nature of the LoRa MAC protocol, with an increase in the number of EDs leading to increased packet collisions. Moreover, when the EDs utilize the highest SF, their ToA values are increased, rendering them more susceptible to interference from other EDs. Therefore, this susceptibility may lead to increased packet collisions, ultimately resulting in a reduction in the accuracy of the analytical model.

Then, the number of EDs is fixed at 160, while the number of GWs varies from 2 to 4.
All the EDs are assigned the largest SF.
As shown in Fig. \ref{MAE_GW}, the MAE is less than 3\% and decreases slightly with the increasing number of GWs.
This is because the increase in the number of deployed GWs reduces the distance between the majority of EDs and their closest GW.
Consequently, this reduction in distance results in shorter ToA values and channel occupation durations, effectively alleviating the negative impact of channel fading on the accuracy of the model.

\begin{table}[h]
	\centering
	\caption{Parameter setting \cite{bor2016lora}}
	\label{threshold}
	\setlength{\tabcolsep}{5mm}{
		\begin{tabular}{cccc}
			\hline  
			& & &\\[-5pt]
			\textbf{Parameter}&$PS^1$&$PS^2$&$PS^3$ \\
			\hline
			& & &\\[-5pt]
			\text{SF}&7&12&12\\
			
			& & &\\[-5pt]
			\text{BW (kHz)}&500&125&125\\
			
			& & &\\[-5pt]
			\text{CR}&4/5&4/8&4/5\\
			\hline
		\end{tabular}\label{PS}}
\end{table} 

Third, we consider three distinct transmission parameter setting policies, $PS^1$, $PS^2$ and $PS^3$ (see Table \ref{PS}), within networks with 3 GWs and 160 EDs.
$PS^1$ utilizes the transmitter settings, leading to the shortest possible ToA of 9.024 ms.
$PS^2$ employs the most robust transmission parameter settings, resulting in the longest possible ToA of 1187.84 ms. Finally, $PS^3$ adopts the parameter settings used by common LoRaWAN deployments \cite{bor2016lora}.
As shown in Fig. \ref{MAE_algorithm}, the MAE of the model is consistently less than 4\%. Nevertheless, the transmission parameter settings in $PS^1$ are obtained with a slightly lower accuracy.
This is because numerous EDs located far from the GW are allocated to the smallest SF, which causes the received signal power at the GW to be lower than the receiver sensitivity threshold, thereby leading to increased packet loss.
Consequently, the model tends to slightly overestimate the PDR.

The entirety of the presented results validate the ability of the proposed analytical model to accurately estimate the PDR regardless of the number of EDs or GWs within the LoRa networks or the transmission parameter settings.

Subsequently, based on this analytical model, the performance of the proposed MMALoRa algorithm in assigning transmission parameters in various LoRa network scenarios is evaluated.
In the subsequent evaluation, the number of CHs is set to $C=4$, and each CH operates with a bandwidth of 125 kHz \cite{yu2022loradar}.

\subsection{Behaviour of the MMALoRa Algorithm}

\begin{figure}[tb]
	\centering
	\subfigure[System EE vs. the number of iterations ($K$=2, $N$=80, 120, 160).]{
		\label{Conv_ED}
		\includegraphics[width=75mm]{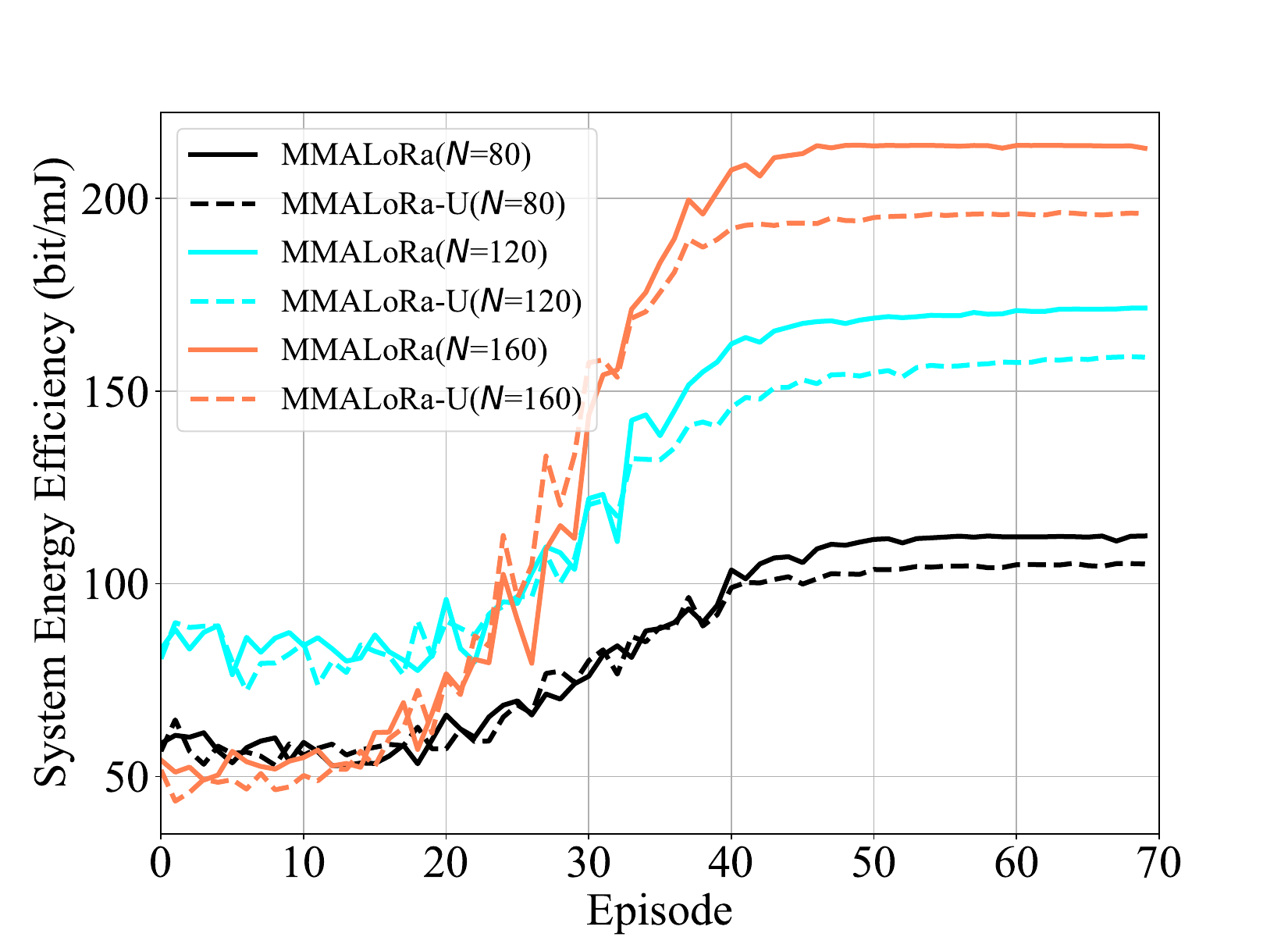}}\\
	\subfigure[The training process of the MMALoRa algorithm ($K$=2, $N$=160).]{
		\label{head_ED}
		\includegraphics[width=75mm]{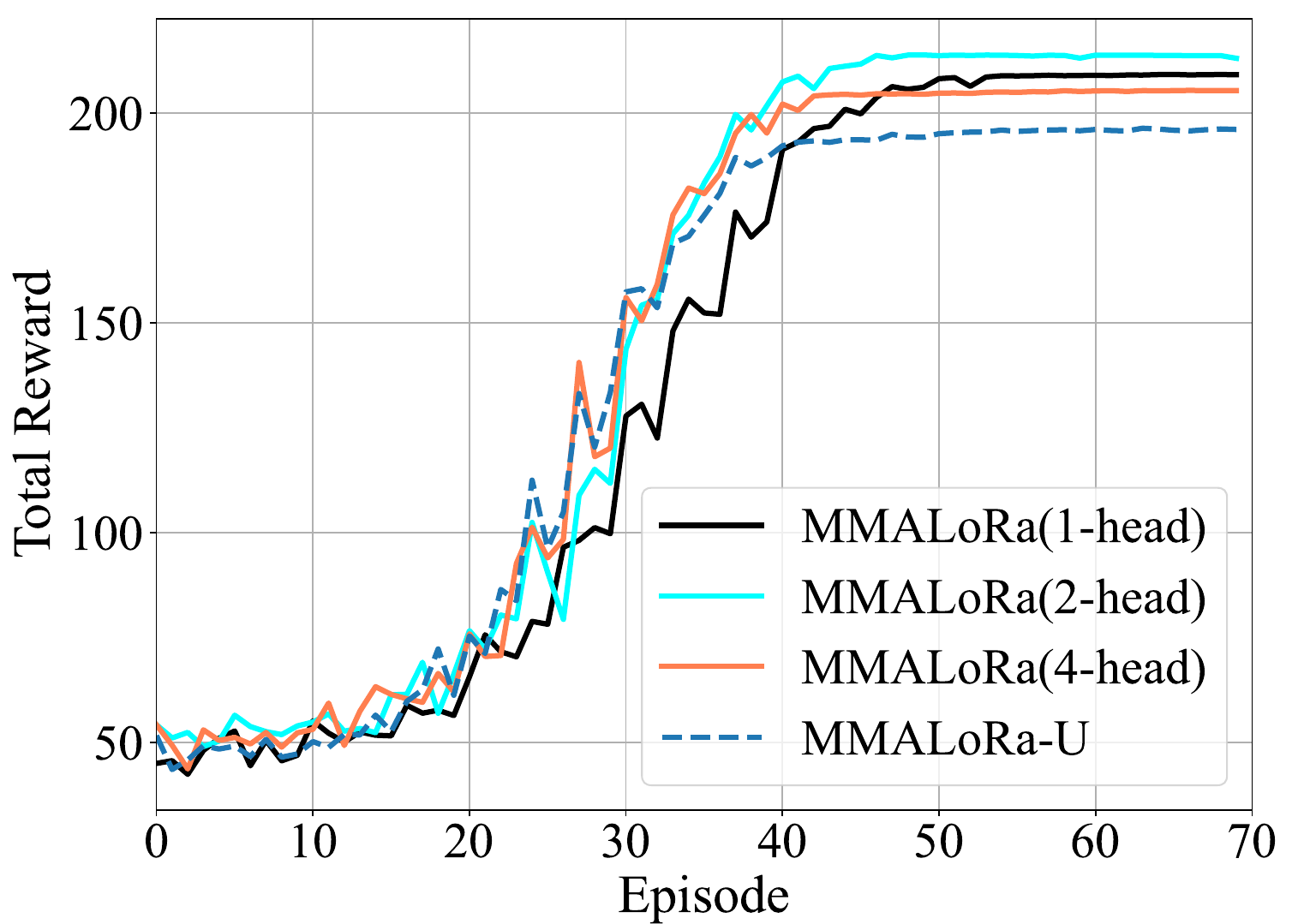}}\\
	\caption{Convergence of the MMALoRa algorithm ($D_{th}$ = 70\%).}
	\label{Convergence}\vspace{-0.2cm}
\end{figure}

To evaluate the convergence of the MMALoRa algorithm for different parameter configurations, we conduct experiments by varying the number of EDs from 80 to 160 while keeping the number of GWs constant at 2. The parameters used for training the MMALoRa algorithm are summarized in Table \ref{train_parameter}. As shown in Fig. \ref{Conv_ED}, the MMALoRa algorithm converges within 50 episodes, with each episode lasting approximately 3 minutes, which is acceptable. Moreover, to evaluate the effectiveness of the attention mechanism, a comparative analysis of the results from the proposed MMALoRa and MMALoRa-U methods is conducted. The MMALoRa-U represents the algorithm with a uniform attention approach in which the attention weights are fixed. This restriction prevents the algorithm from focusing on specific EDs. As shown in Fig. \ref{Conv_ED}, the system EE with the MMALoRa algorithm is better than that with the MMALoRa-U algorithm, and the difference in the system EEs of the MMALoRa and MMALoRa-U algorithms increases with increasing number of EDs. 
This is not surprising because the MMALoRa-U algorithm focuses on all agents equally, which prevents each agent from obtaining valuable information from other agents during the training process.
Consequently, MMALoRa-U fails to effectively mitigate the cochannel interference that arises with an increasing number of EDs.

Then, the number of attention heads is varied to examine the effect of the attention mechanism on the performance of the MMALoRa algorithm. The numbers of GWs and EDs are set to 2 and 160, respectively. As shown in Fig. \ref{head_ED}, the total reward achieved by the 4-head MMALoRa algorithm is lower than the rewards achieved by the 1-head MMALoRa and 2-head MMALoRa algorithms. This is not surprising because the increased number of attention heads complicates the algorithm, which can potentially lead to overfitting in instances where the data available in the experience replay buffer are insufficient for training.

\subsection{Comparison with Existing Algorithms}

Three baseline schemes are used to analyse the performance of the MMALoRa algorithm.
\begin{itemize}
	\item \textbf{ADR}: The ADR algorithm adaptively selects SF and TP values for the EDs according to the SNR and ensures that the SNR is higher than the demodulation floor \cite{garlisi2020capture}.
	
	\item \textbf{EF-LoRa}: EF-LoRa uses a greedy algorithm that aims at maximizing the minimum EE in LoRa networks \cite{gao2019towards}.
	
	\item \textbf{RCST}: RCST employs a random selection mechanism to determine the CHs, SFs and TPs.
\end{itemize}

\begin{figure}[tb]
	\centering
	\subfigure[Average PDR vs. different number of EDs.]{
		\label{pdr_ED}
		\includegraphics[width=75mm]{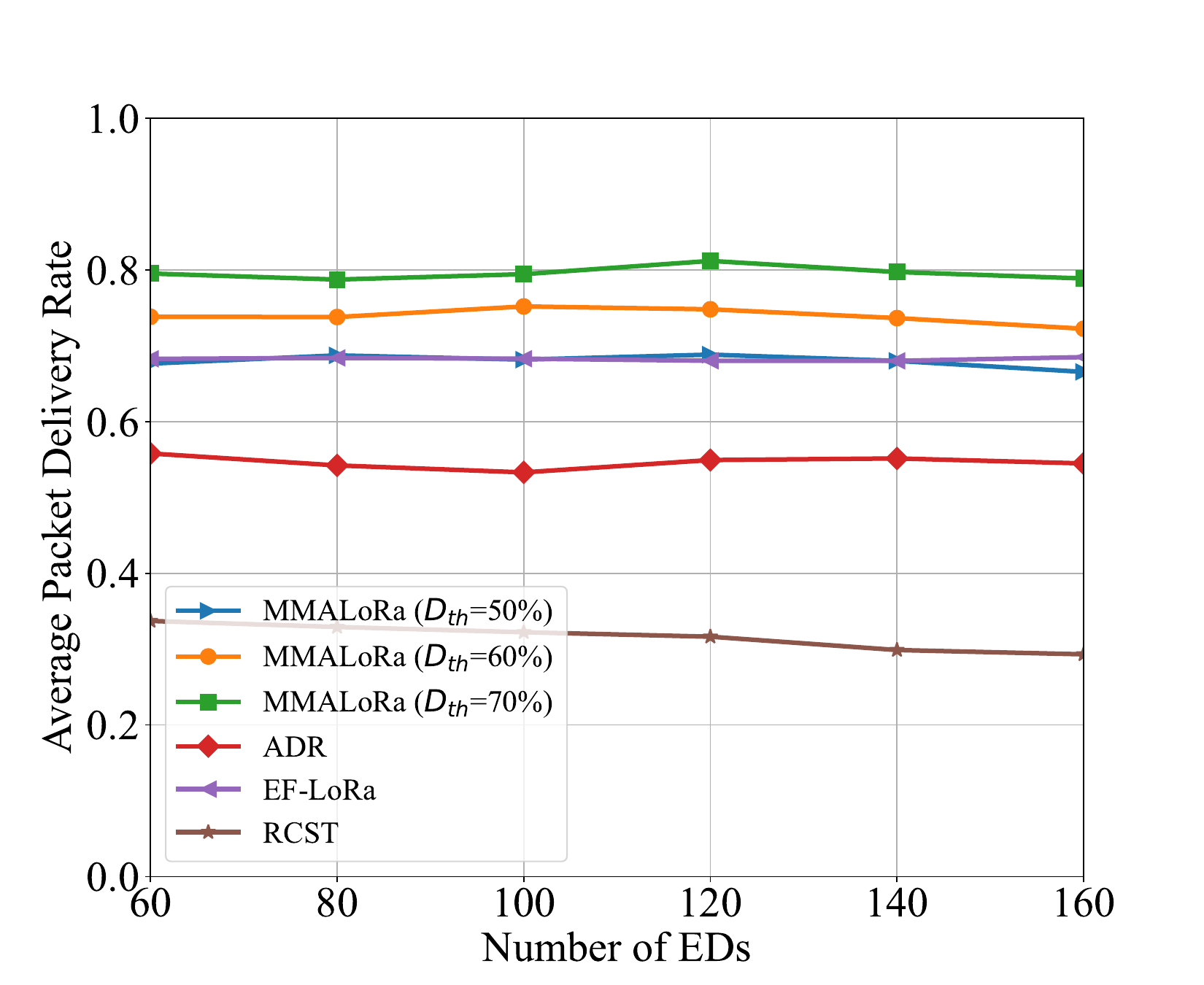}}\\
	\subfigure[System EE vs. different number of EDs.]{
		\label{EE_ED}
		\includegraphics[width=75mm]{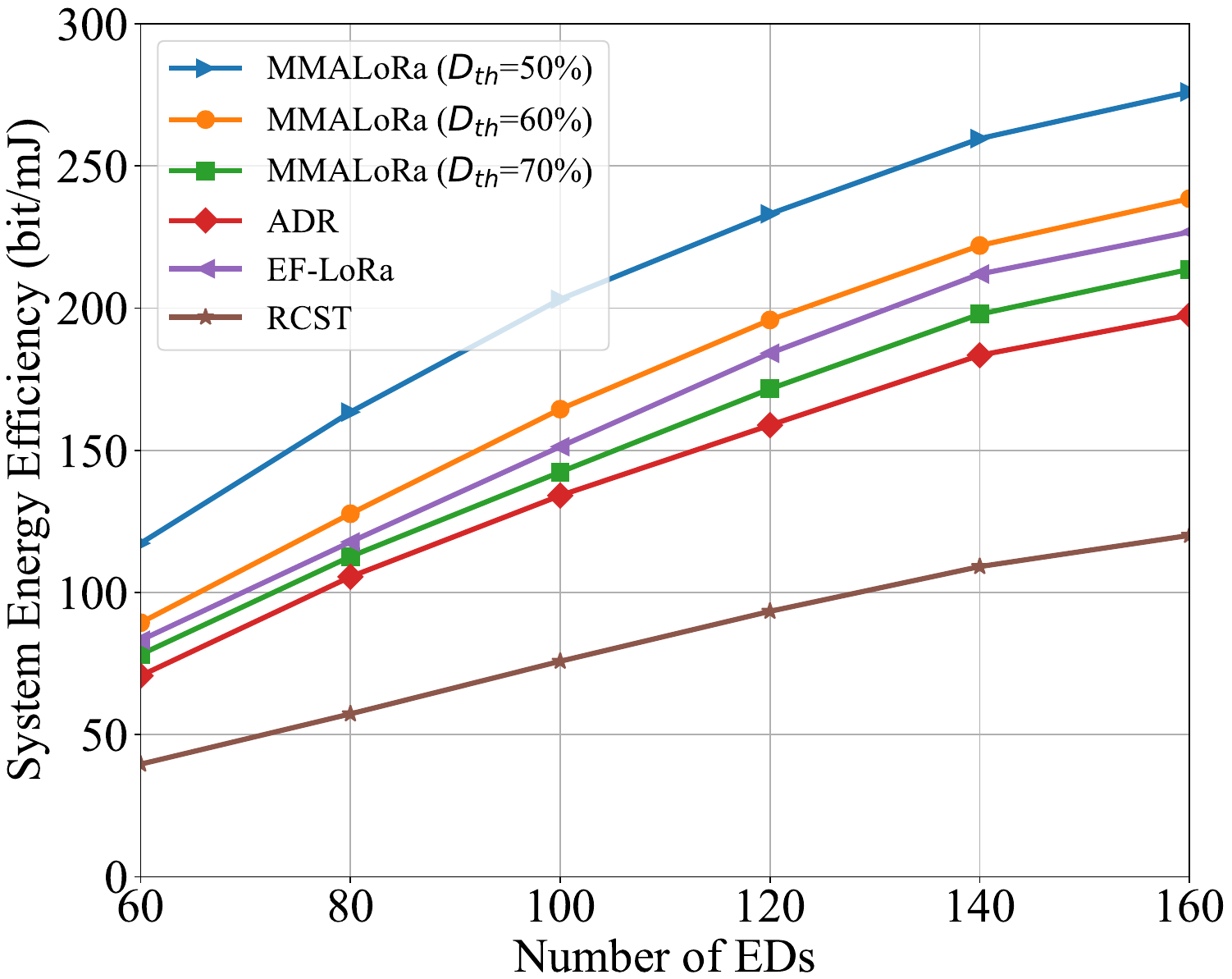}}\\
	\caption{Comparison of the average PDR and system EE results achieved by different algorithms under different numbers of EDs ($K$=3).}
	\label{ED}\vspace{-0.2cm}
\end{figure}

\begin{figure}[tb]
	\centering
	\subfigure[Average PDR vs. different number of GWs.]{
		\label{pdr_GW}
		\includegraphics[width=75mm]{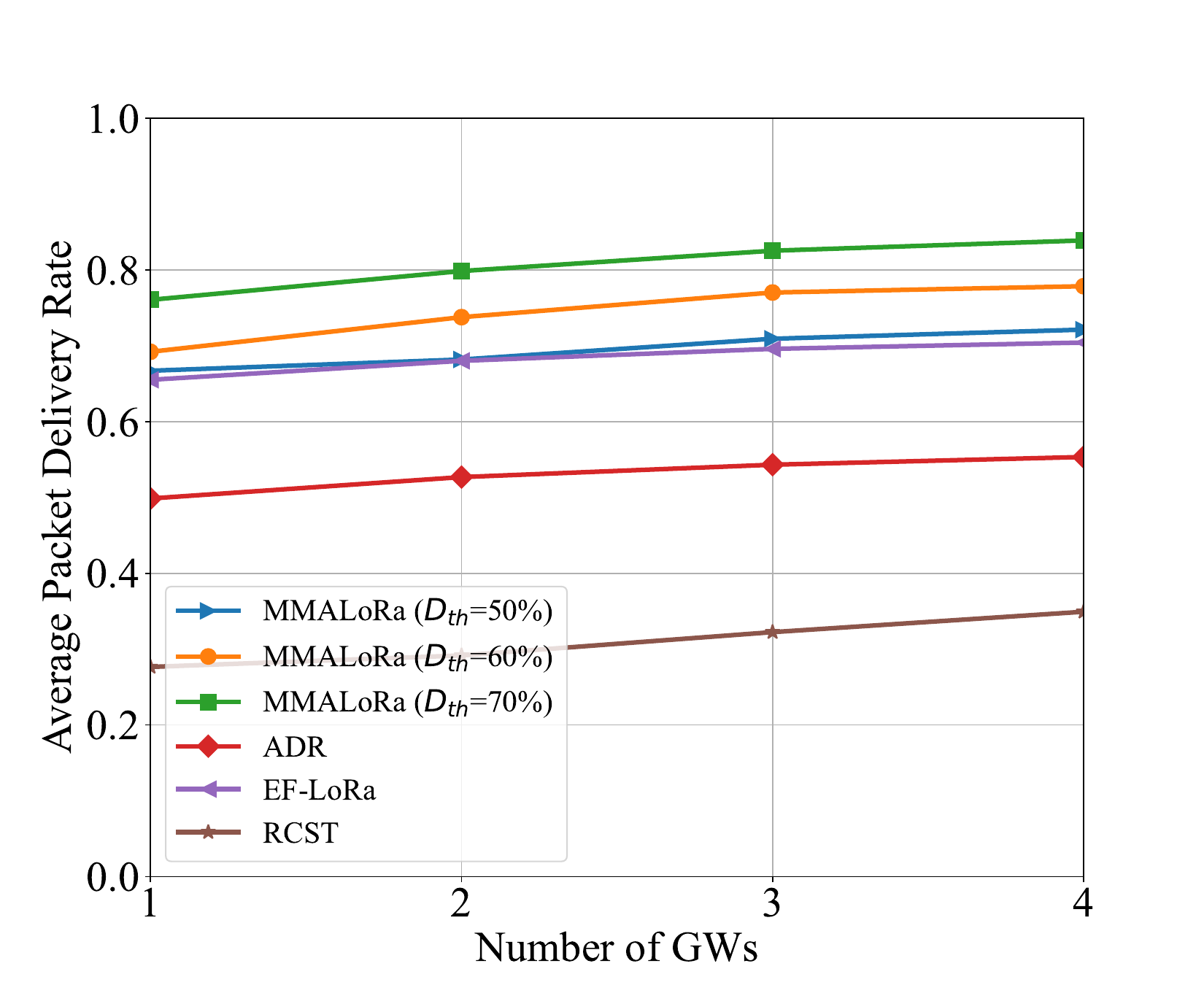}}\\
	\subfigure[System EE vs. different number of GWs.]{
		\label{EE_GW}
		\includegraphics[width=75mm]{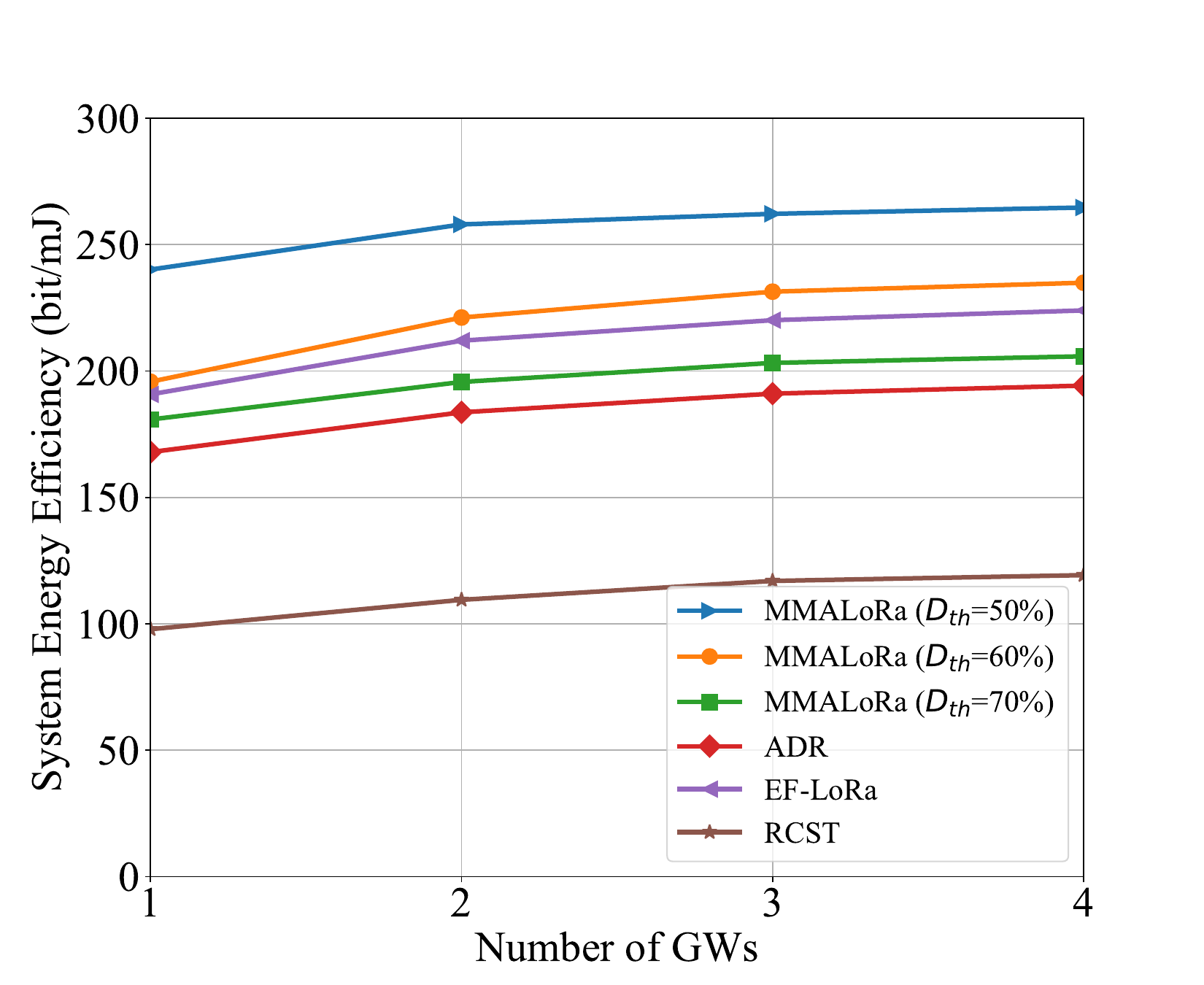}}\\
	\caption{Comparison of the average PDR and system EE results achieved by different algorithms under different numbers of GWs ($N$=140).}
	\label{GW}\vspace{-0.2cm}
\end{figure}

The PDR constraint $D_{\text{th}}$ is varied among values of 50\%, 60\%, and 70\% to investigate its effect on the performance of the MMALoRa algorithm.

We first conduct a comparative analysis of the average PDR and system EE results obtained by employing various algorithms with different numbers of EDs. The number of GWs is fixed at 3, while the number of EDs varies from 60 to 160.
As shown in Fig. \ref{ED}, the MMALoRa algorithm can optimize the system EE while satisfying varying PDR constraints. Specifically, as depicted in Fig. \ref{pdr_ED}, when subjected to PDR constraints of $D_{\text{th}}$ at 50\%, 60\%, and 70\%, the MMALoRa algorithm achieves average PDRs of approximately 70\%, 75\%, and 80\%, respectively.
Fig. \ref{EE_ED} shows the system EE versus the number of EDs. Furthermore, the performance gap between the MMALoRa algorithm and the other algorithms escalates with increasing number of EDs.
This is because with an increasing number of EDs, the MMALoRa algorithm can intelligently allocate transmission parameters to mitigate cochannel interference.
The EF-LoRa algorithm obtains limited performance because it is designed to address a max-min problem to achieve EE fairness among the EDs.
On the other hand, the ADR algorithm tends to select larger SFs and higher TPs to maintain the SNR above the demodulation threshold, resulting in limited effectiveness in mitigating interference.
However, as the PDR constraint increases, the system EE calculated by the MMALoRa algorithm decreases. Notably, when the PDR constraint $D_{\text{th}}$ is set to 70\%, the MMALoRa algorithm demonstrates limited performance in terms of the system EE.
This is because the MMALoRa algorithm tends to allocate higher SFs and TPs to EDs to ensure reliable transmission, which consequently leads to an increase in energy consumption.
Next, we compare the performance of different algorithms with varying numbers of GWs. The number of GWs is varied from 1 to 4, and the number of EDs is set to 140.
As shown in Fig. \ref{GW}, the system EE and average PDR improve as the number of GWs increases.
This is because the increasing number of GWs may reduce the distances between EDs and their nearest GWs. Consequently, EDs can utilize lower SFs and TPs, thereby achieving lower energy consumption and shorter ToA values. The reduced ToA also diminishes the probability of packet collisions due to the interference from other EDs.

\section{Conclusion}

In this paper, we investigate multi-GW LoRa networks and propose an analytical model to calculate the system EE while fully considering duty cycling, quasi-orthogonality and capture effects.
The simulation results validate that the proposed analytical model accurately assesses the system EE of LoRa networks under different settings according to comparisons with the system EE obtained by the open-source LoRasim simulator.
On the basis of the analytical model, we investigate the problem of joint CH, SF and TP allocation to optimize the system EE. Owing to the NP-hardness of the problem, we decomposed the original problem into two subproblems: CH assignment and SF/TP assignment. Then, a two-stage optimization framework, MMALoRa, is proposed to solve the two subproblems.
The simulation results demonstrate that the MMALoRa algorithm rapidly converges under various parameter configurations, significantly enhancing the system EE while ensuring the PDR constraint in comparison to the three baseline algorithms.







\end{document}